\documentclass[final,3p]{elsarticle}

\usepackage[colorlinks=true,breaklinks=true,pdftex]{hyperref}

\usepackage{amssymb,amsmath,graphicx,bm,siunitx}
\usepackage{booktabs,threeparttable,diagbox,multirow,hyperref,float}
\usepackage[ruled]{algorithm2e} 
\usepackage[labelformat=simple]{subcaption}

\UseRawInputEncoding

\journal{Computer aided geometric design}

\newcommand{\mpp}{\textbf{\emph{p}}}
\newcommand{\mc}{\textbf{\emph{c}}}

\newcommand{\boldcal}[1]{\bm{\mathcal{#1}}}

\biboptions{authoryear}
\begin{document}

\begin{frontmatter}

\title{$p\kappa$-Curves: Interpolatory curves with curvature approximating a parabola}



\author[author1,author2]{Zhihao Wang}
\author[author1,author2]{Juan Cao\corref{mycorrespondingauthor}}
\cortext[mycorrespondingauthor]{Corresponding author}
\ead{juancao@xmu.edu.cn}
\author[author1,author2]{Tuan Guan}
\author[author6]{Zhonggui Chen}
\author[author8]{Yongjie Jessica Zhang}
\address[author1]{School of Mathematical Sciences, Xiamen University, Xiamen, Fujian, 361005, China}
\address[author2]{Fujian Provincial Key Laboratory of Mathematical Modeling and High-Performance Scientific Computation,\\ Xiamen University, Xiamen, Fujian, 361005, China}
\address[author6]{School of Informatics, Xiamen University, Xiamen, Fujian, 361005, China}
\address[author8]{Department of Mechanical Engineering, Carnegie Mellon University, Pittsburgh, PA 15213, USA}

\begin{abstract}
This paper introduces a novel class of fair and interpolatory curves called $p\kappa$-curves. These curves are comprised of smoothly stitched B\'ezier curve segments, where the curvature distribution of each segment is made to closely resemble a parabola, resulting in an aesthetically pleasing shape. Moreover, each segment passes through an interpolated point at a parameter where the parabola has an extremum, encouraging the alignment of interpolated points with curvature extrema. To achieve these properties, we tailor an energy function that guides the optimization process to obtain the desired curve characteristics. Additionally, we develop an efficient algorithm and an initialization method, enabling interactive modeling of the $p\kappa$-curves without the need for global optimization. We provide various examples and comparisons with existing state-of-the-art methods to demonstrate the curve modeling capabilities and visually pleasing appearance of $p\kappa$-curves.

\end{abstract}

\begin{keyword}
Fairness \sep interpolation \sep continuity  \sep smoothness \sep locality
\end{keyword}

\end{frontmatter}


\section{Introduction}
Interpolatory curves play a crucial role in computer-aided geometric design (CAGD) and find extensive applications in various fields such as industrial design, art design, shape representation, animation, and more~\citep{farin2002curves}. Piecewise parametric curves are commonly employed to design complex shapes due to their flexibility ~\citep{piegl:1996:SSBM}. As a result, the construction of piecewise interpolation curves has become a popular research topic. In addition to the fundamental requirement of interpolation, there is often a need to satisfy additional geometric properties such as smoothness, locality, fairness, robustness, roundness, extensionality, and more~\citep{levien2009spiral}. Among these properties, smoothness, fairness, and locality are considered key geometric properties that are highly desired in practical curve design~\citep{binninger2022smooth,Wang:2010:CMAME}.

In curve design, achieving smooth and visually appealing transitions between curve segments is important. The smoothness is typically evaluated through two types of continuity: parametric continuity ($C^n$-continuity) and geometric continuity ($G^n$-continuity). $C^n$-continuity requires the matching of parametric derivatives up to order $n$ at joints where the curve segments are stitched together. On the other hand, $G^n$-continuity demands the existence of some reparameterization of the segments, resulting in $C^n$-continuity at the joints. The choice of continuity order depends on the specific application. In many cases, achieving second-order continuity ($C^2$- or $G^2$-continuity) is desirable in curve design with multiple segments due to its practical and aesthetic benefits~\citep{Sederberg:2012}.

The concept of fairness in curve design is subjective and lacks a precise mathematical criterion for evaluation. However, studies in psychology have shown that the fairness of a curve is strongly correlated with its curvature~\citep{attneave1954some,levien2009spiral}. Curves that exhibit smooth curvature variation and have fewer curvature monotonic intervals are commonly associated with a perception of having a more fairing shape~\citep{farin2002curves}. Previous methods have commonly emphasized the necessity of maintaining monotonic curvature variation on each curve segment to enhance the smoothness and aesthetic quality of curves~\citep{mineur:1998:CAGD,cao:2008:CAGD}. However, in curve construction with multiple segments, achieving both smoothness and fairness in higher-order curves poses challenges due to the intricate task of preserving curvature monotonicity while handling non-unique curvature extreme points. In addressing the fairing problem, some other approaches formulate it as an energy optimization problem~\citep{zhang:2001:CAD,levien2009interpolating,jiang:2023:TVC}. However, these approaches do not explicitly consider curvature features such as monotonicity and the position of local curvature extremes.

The locality property of curves means that changes in the position of an interpolated point only affect its adjacent region of the curve. This enables easier and more intuitive local adjustments to the curve shape, allowing users to fine-tune specific regions of interest. Additionally, it improves computational efficiency by limiting calculations to the affected local portions of the curve, reducing complexity and speeding up the curve editing process. In general, there is a tradeoff between achieving locality and higher-order continuity, i.e. curves with higher-order continuity usually have weaker locality~\citep{levien2009interpolating}. In particular, methods that construct curves relying on global optimization techniques typically do not exhibit locality.

To design interpolatory curves with complex shapes, geometric primitives such as (rational) B\'ezier curves~\citep{yan2017k,yan2019circle}, trigonometric blending curves~\citep{yuksel2020class}, clothoids and straight lines~\citep{binninger2022smooth} are commonly used. 
In this paper, we propose an efficient method for constructing 2D interpolatory curves that satisfy three specific properties (smoothness, fairness, and locality), using quartic and quintic B\'{e}zier curves as the geometric primitives. Different from the traditional criterion of monotonic curvature variation, our approach focuses on encouraging each segment to closely approximate a parabolic shape in terms of curvature variation. Additionally, we align the interpolation point with the extreme value of the approximated parabola. Compared to the previous energy optimization approaches, the resulting curve segments tend to have a maximum of two monotonic intervals in terms of curvature variation. This approach leads to fairing interpolatory curve designs. We introduce ``$p\kappa$-curves'' to refer to these constructed curves, which consist of segments with approximate parabolic curvature variations. Our specific contributions are as follows:
\begin{description}
  \item[$\bullet$] We propose an optimization-based method for constructing fair and interpolatory $p\kappa$-curves using B\'ezier segments. We have tailored an energy function to guide the optimization process and obtain these curves. The resulting $p\kappa$-curves exhibit several desirable properties, such as a curvature distribution in each segment that closely resembles a parabola, interpolated points closely aligning with curvature extrema, up to $C^2/G^2$ smoothness, and a strong locality property.
  \item[$\bullet$] We develop an efficient algorithm for locally solving the optimization problem. This algorithm is equipped with an efficient initialization method, enabling interactive design and local editing. Experimental results demonstrate the capability of our method in diverse shape design, producing aesthetically pleasing curves.
\end{description}

The paper is organized as follows: Section~\ref{sec:related_work} provides a review of the related works on interpolatory curves. Section~\ref{sec:preliminary} offers a brief introduction to the definition, calculation, properties, and notation of B\'{e}zier curves. The optimization problem for the construction of $p\kappa$-curves is presented in Section~\ref{sec:method}. The optimization method for interactive interpolatory curve design is given in Section~\ref{sec:interaction}. Section~\ref{sec:experiments} conducts an ablation study, presents the experimental results, and provides a comparison with state-of-the-art methods.  Finally, Section~\ref{sec:conclusion} concludes the paper.

\section{Related work} \label{sec:related_work}

Extensive research has focused on constructing interpolatory curves using multiple geometric primitives. According to the primitives used, these methods can be categorized into three groups: polynomial curves (including their rational forms), logarithmic aesthetic curves, and blending of different primitives. For a comprehensive survey on this topic, please refer to~\citep{hoschek1993fundamentals,levien2009spiral,miura2014aesthetic}.

\subsection{Polynomial curves}

Polynomial representations like B\'{e}zier, B-spline, and NURBS have found widespread use in constructing free-form curves within Computer-Aided Geometric Design (CAGD)~\citep{piegl:1996:SSBM}. These representations have gained popularity due to their simplicity and intuitive controls. Researchers have proposed different types of smooth interpolation curves, such as Catmull-Rom splines~\citep{catmull1974class,barry1988recursive}, $G^2$ cubic splines~\citep{farin2002curves} and subdivision curves~\citep{deslauriers1989symmetric}. However, the main focus of these earlier works has been on achieving smoothness and accurate interpolation, with less consideration given to curve fairness, resulting in unexpected geometric feature points (cusps, inflections, and loops).

$\kappa$-curves, introduced by \cite{yan2017k}, offer improved control over the curvature distribution of interpolatory curves. These curves are constructed by connecting a series of quadratic B\'{e}zier curve segments, where each segment is specifically designed to have its unique curvature maximum aligned with a given point. This concept was later extended through the utilization of rational quadratic B\'{e}zier curves, enabling accurate representation of circles, as well as other elliptical or hyperbolic shapes~\citep{yan2019circle}.
$\epsilon\kappa$-curves~\citep{miura2022epsilon} are another extension of the $\kappa$-curves that employ cubic B\'{e}zier curves. By utilizing higher-degree primitives, $\epsilon\kappa$-curves introduce additional degrees of freedom (DOFs), enabling precise control over the magnitudes of local maximum curvature. Due to their reliance on global optimization techniques, $\kappa$-curves and their generalizations lack locality. Additionally, because the number of DOFs is limited, neither $\kappa$-curves nor their generalizations can guarantee global second-order continuity. The feature points controlled interpolatory curves (FPC-curves) also utilize cubic B\'{e}zier curves, with a primary objective of efficiently and precisely controlling the location and type of geometric feature points~\citep{chen2019interpolatory}. FPC-curves exhibit good local properties, but at joints where there is a sign change in curvature, the continuity is reduced to $G^1$. Directly utilizing higher-order polynomial primitives provides increased DOFs for achieving higher-order continuity. However, it also increases the difficulty in controlling curvature variation due to the presence of multiple curvature extrema within each high-order segment.

Alternatively, several research works have focused on obtaining fair curves by minimizing specific energy functions tailored for polynomial representation. These energy functions include stretch energy~\citep{xu:2011:SCIS}, strain energy~\citep{chan:2011:matematika,xu:2011:SCIS}, jerk energy~\citep{ericskin:2016:msae,xu:2011:SCIS}, bending energy~\citep{johnson:2020:CAGD}, and others. These energy functions are generally defined globally, which would lead to curves that lack locality.  In a recent work,~\cite{jiang:2023:TVC} employed the progressive iterative approximation method for fairing curve construction, allowing for local improvements. However, none of these methods directly considers the variation of curvature or the positions of local curvature extrema.

\subsection{Log-aesthetic curves}
Log-aesthetic curves (LACs) are visually pleasing curves with linear curvature and torsion graphs, resulting in a monotonically varying curvature~\citep{miura:2005:ICHC}. These curves are highly valued in aesthetic design, and previous research has developed smooth interpolation curves based on log-aesthetic curves~\citep{miura:2013:CADA,Wang:2021:CADA}. LACs include well-known curves such as the logarithmic spiral, clothoid, and involute of a circle. Among them, the clothoid (Euler spiral) has a linear variation of curvature with respect to arc length, making it suitable for achieving local maximum curvature at interpolated points.
Extensive research has been conducted on constructing clothoid-based interpolated curves using global optimization methods, such as~\citep{havemann2013curvature,bertolazzi:2018:MMAS}. More recently, \cite{binninger2022smooth} proposed a planar curve comprised of piecewise clothoids and straight lines, allowing local shape control by adjusting the position of interpolated points and re-estimating the associated curvatures. It is worth noting that LACs generally lack a closed-form representation as they are defined via Fresnel integrals~\citep{Olver:2010:CU}. Hence, numerical methods are commonly employed for curve construction and computation of values and derivatives.


\subsection{Blending curves}
Blending curves are constructed by combining different types of geometric primitives, such as (rational) B\'{e}zier curves, (rational) B-spline curves, line segments, parabolas, and circle arcs. In early works on blending curve construction, the main emphasis was on achieving $G^1$-continuity in interpolation, for example, by blending circular arcs and straight lines~\citep{wenz:1996:CAGD} and linearly blending weight functions~\citep{overhauser:1968,wenz:1996:CAGD,pobegailo:1992:TVC}. Subsequently, considerable research efforts have focused on achieving $C^2$-continuity. \cite{wiltsche:2005:JGG} introduced polynomial functions for blending two arbitrarily parametric curves. Other approaches utilize B-spline basis functions to blend Lagrange~\citep{wiltsche:2005:JGG} or Hermite interpolants~\citep{gfrerrer:2001:CAGD}.
Polynomial blending functions, as presented in~\citep{pobegailo:2013:IJCGA}, offer the ability to construct curves with arbitrarily high orders of continuity. Trigonometric blending functions have also been applied in interpolating circular arcs, as demonstrated in~\citep{szilvasi:2000:CAGD, sequin:2005:CAD}. In~\citep{sun:2009:CG}, a similar formulation using rational quadratic B\'ezier interpolation was introduced, allowing for the creation of conic sections. Recently, \citep{yuksel2020class} introduced a method that utilizes trigonometric blending of hybrid circular-elliptical interpolations. This method constructs curves that are free from cusps or self-intersections between two consecutive interpolated points. While all these blending curves exhibit locality and computational efficiency, they do not take into account the variation of curvature and the position of curvature extrema.

\section{Preliminary\label{sec:preliminary}}
In this section, we provide a brief introduction to the definitions, calculations, properties, and notations of B\'{e}zier curves, which are relevant to our study.

\subsection{B\'ezier curves \label{sec:bezier}}
In this paper, we utilize B\'{e}zier curves as the geometric primitives for constructing our interpolatory curves. Assume an interpolatory curve  $\{\boldcal{P}_{i,k}(t)\}_{i=0}^m$ consists of $m+1$ degree-$k$ B\'{e}zier segments. The $i$-th segment of the parametric curve sequence can be defined as follows:
\begin{equation}\label{eq:bezier}\boldcal{P}_{i,k}(t) = \sum_{j=0}^k \mc_{i,j} B_j^k(t), \quad t\in[0,1],
\end{equation}
where $\mc_{i,j}\in \mathbb{R}^2$ are the control points and $B_j^k(t)$ denotes the degree-$k$ Bernstein base function given by
$$B_j^k(t)=\frac{k!}{j!(k-j)!}t^j(1-t)^{k-j}.$$ The polyline formed by sequentially connecting the control points $\mc_{i,j}$ $(j=0,...,k)$ is referred to as the control polygon of the B\'{e}zier curve $\boldcal{P}_{i,k}(t)$.   The curvature of $\boldcal{P}_{i,k}(t)$ can be computed as follows:
\begin{align}
        \kappa_i(t) =  \frac{\det ( \boldcal{P}^{'}_{i,k}(t),  \boldcal{P}^{''}_{i,k}(t))}{||\boldcal{P}^{'}_{i,k}(t)||^3},
    \end{align}
where $\mathrm{det}(\cdot)$ represents the determinant of a matrix, $\cdot^{'}$ and $\cdot^{''}$ denotes the first and second derivative operations, and $||\cdot||$ denotes the length of a vector.

\subsection{Degree elevation and subdivision }

B\'{e}zier curves have found extensive applications in CAGD, owing to the development of various effective algorithms~\citep{farin2002curves}. In this paper, we adopt the degree elevation algorithm and subdivision algorithm, which we briefly review here.

\begin{figure}
  \centering
  \begin{subfigure}[b]{0.32\textwidth}
  \centering
        \includegraphics[width = 1.0 \textwidth]{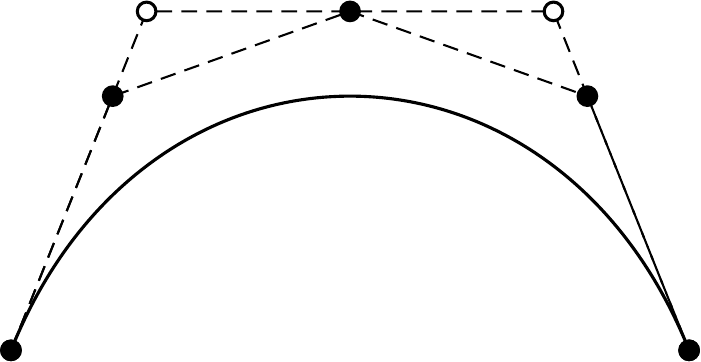}\caption{}
  \end{subfigure} \qquad
  \begin{subfigure}[b]{0.32\textwidth}
  \centering
        \includegraphics[width = 1.0 \textwidth]{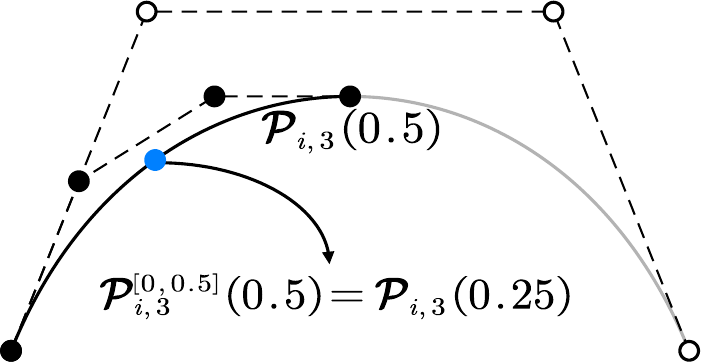}\caption{}
  \end{subfigure}
  \caption{B\'{e}zier curve degree elevation and subdivision. (a) Degree elevation of a cubic B\'{e}zier curve to quartic, and (b) subdivision of B\'{e}zier curve $\boldcal{P}_{i,3}(t)$ at $t=z=0.5$ and the left segment $\boldcal{P}^{[0,0.5]}_{i,3}(\bar{t}) = \boldcal{P}_{i,3}(\bar{t}/2),\bar{t} \in [0,1]$. Original control points are denoted by hollow  points, while control points after the degree elevation and subdivision operations are indicated by solid points.} \label{fig:subdivision}
\end{figure}

\textbf{Degree elevation.} The degree elevation algorithm is used to calculate the control points $\bar{\mc}_{i,j}$ of a degree-($k+1$) B\'{e}zier curve
$$\boldcal{P}_{i,k+1}(t) = \sum\limits_{j=0}^{k+1} \bar{\mc}_{i,j} B_j^k(t), \quad t\in[0,1],$$
that is geometrically and parametrically equivalent to a given degree-$k$ B\'{e}zier curve $\boldcal{P}_{i,k}(t)$ defined in Eq.~(\ref{eq:bezier}), where the control points are:
$$\bar{\mc}_{i,0} = \mc_{i,0}; \quad \bar{\mc}_{i,l} = \frac{l}{k+1}\mc_{i,l-1} + \left(1 - \frac{l}{k+1}\right)\mc_{i,l},~l=1,\ldots,k;  \quad \bar{\mc}_{i,k+1} = \mc_{i,k}.$$
In Fig.~\ref{fig:subdivision}(a), we present an example of a curve, showing the control points both before and after the degree elevation process.

\textbf{Subdivision.} The subdivision algorithm can be used to split a given curve defined in Eq.~(\ref{eq:bezier}) at parameter $t=z$ $(z\in(0, 1))$ into two curve segments $\boldcal{P}^{[0,z]}_{i,k}(\bar{t})$ and $\boldcal{P}^{[z,1]}_{i,k}(\bar{t})$, each of which is still a B\'{e}zier curve. In particular,
the segment of the curve on $[0,z]$ is geometrically equivalent to a degree-$k$ B\'{e}zier curve
$$ \boldcal{P}^{[0,z]}_{i,k}(\bar{t}) = \sum\limits_{j=0}^{k} \bar{\mc}_{i,j} B_j^k(\bar{t}), \quad \bar{t}\in[0,1], $$
where the control points can be computed as
$$
\begin{bmatrix}
\bar{\mc}_{i,0} \\
\bar{\mc}_{i,1} \\
\cdots \\
\bar{\mc}_{i,k}
\end{bmatrix} =
\begin{pmatrix}
1, &0, &\cdots &0 \\
1-z, &z, &\cdots &0 \\
& \vdots & & \\
(1-z)^k, &\frac{k(k-1)}{2}(1-z)^{(k-1)}z, &\cdots &z^k
\end{pmatrix}
\begin{bmatrix}
\mc_{i,0} \\
\mc_{i,1} \\
\cdots \\
\mc_{i,k}
\end{bmatrix}.
$$
A point on the original curve with a parameter value $t \in [0, z]$ corresponds to a new parameter value $\bar{t} = t/z$ on the new representation, specifically $\boldcal{P}^{[0,z]}_{i,k}(\bar{t}) = \boldcal{P}_{i,k}(t)$. In Fig.~\ref{fig:subdivision}(b), we demonstrate B\'{e}zier subdivision, splitting the original curve at $t=0.5$, where we have $\boldcal{P}^{[0,z]}_{i,k}(\bar{t}) = \boldcal{P}_{i,k}(\bar{t}/2)$. By applying the symmetry property of the B\'ezier curve, we can also determine the control points for the segment $\boldcal{P}^{[z,1]}_{i,k}(\bar{t})$ on $[z, 1]$.

\subsection{Continuity constrains\label{sec:continuity}}
\begin{figure}
  \centering
  \begin{subfigure}[b]{0.55\textwidth}
  \centering
        \includegraphics[width = 1.0 \textwidth]{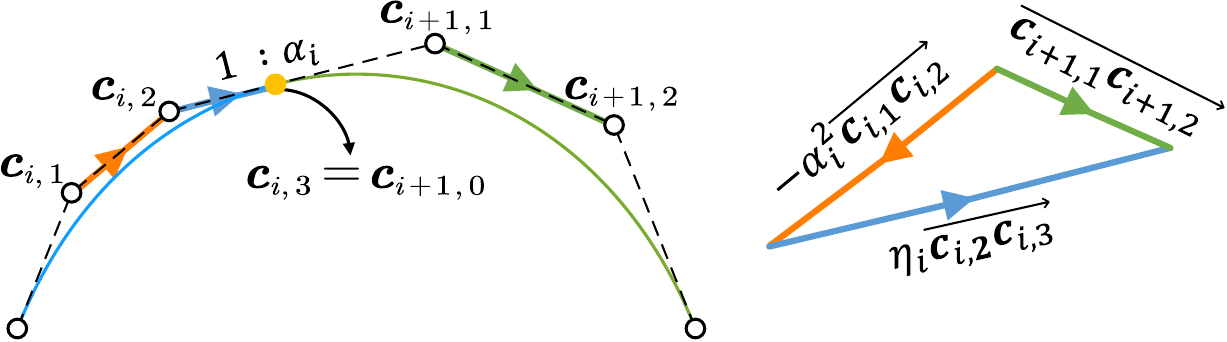}\caption{}
  \end{subfigure} \hspace{1cm}
    \begin{subfigure}[b]{0.32\textwidth}
  \centering
        \includegraphics[width = 1.0 \textwidth]{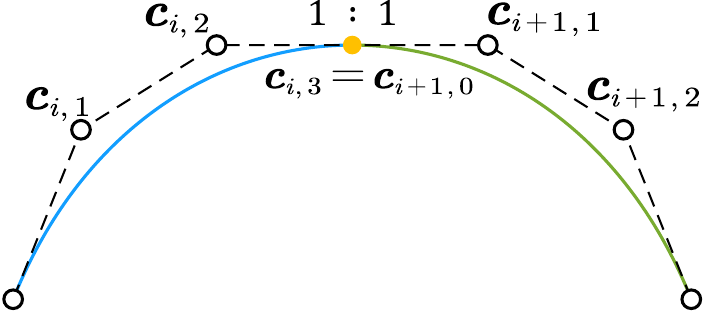}\caption{}
  \end{subfigure}
  \caption{$G^2$- and $C^2$-continuity constraints for B\'{e}zier Curves. The two curves are colored in blue and green, respectively. (a) $G^2$-continuity, where the control legs of the two curves at the joint (yellow point) are collinear with a length ratio of $1:\alpha_i$. Additionally, the last two control legs of the left curve, scaled by factors $-\alpha_i^2$ and $\eta_i$ respectively, form a triangle with the second control leg of the right curve. (b) $C^2$-continuity, a specific case of (a) with $\alpha_i=1$ and $\eta_i=2$.} \label{fig:continuity}
\end{figure}
When constructing interpolatory curves using B\'{e}zier curves, achieving higher-order continuity at the joining points is crucial as it greatly impacts the aesthetic quality of the curve shape. There are two types of continuity in B\'{e}zier curves: parametric continuity ($C^n$) and geometric continuity ($G^n$). Parametric continuity emphasizes smoothness in the curve's parameterization, while geometric continuity focuses on visual smoothness in the curve's shape. Given the adjacent B\'ezier curves ${\boldcal{P}_{i,k}}(t),{\boldcal{P}_{i+1,k}}(t)$, the constraints for ensuring $C^n$ and $G^n$-continuity at the joint $\boldcal{P}_{i,k}(1)=\boldcal{P}_{i+1,k}(0)$ are denoted as  $C^{C,n}_i$ and $C^{G,n}_i$, respectively. Specifically, the constraints for achieving first-order and second-order continuities in the 2D plane are as follows~\citep{Sederberg:2012}:
\begin{description}
\item[$\bullet$] $C^{C,1}_i$: $\mc_{i, k}=\mc_{i+1,0}$ and $\mc_{i, k}-\mc_{i, k-1}=\mc_{i+1,1} - \mc_{i+1,0}$.
\item[$\bullet$] $C^{C,2}_i$: $C^{C,1}_i$ and $\mc_{i, k-2} - 2\mc_{i, k-1}=\mc_{i+1,2} - 2\mc_{i+1,1}$.
\item[$\bullet$] $C^{G,1}_i$: $\mc_{i, k}=\mc_{i+1,0}$ and $\exists~\alpha_i>0$ s.t. $\alpha_i(\mc_{i, k}-\mc_{i, k-1})=\mc_{i+1,1} - \mc_{i+1,0}.$
\item[$\bullet$] $C^{G,2}_i$: $C^{G,1}_i$ and $\exists~\eta_i\in \mathbb{R}$ $s.t.$ $ -\alpha_i^2\left(\mc_{i,k-1}-\mc_{i,k-2}\right)+\eta_i\left(\mc_{i,k}-\mc_{i,k-1}\right)=\mc_{i+1,2}-\mc_{i+1,1}.$
\end{description}
The constraints for achieving $G^2$- and $C^2$-continuity between two B\'{e}zier curves at their joint are shown in Fig.~\ref{fig:continuity}. For a more in-depth discussion on parametric and geometric continuity, we recommend referring to~\citep{farin2002curves}.

\section{Problem statement} \label{sec:method}
In this section, we define the optimization problem for constructing $p\kappa$-curves, including the key objectives and constraints.

\subsection{Constraints}

\begin{figure}
  \centering
  \begin{subfigure}[b]{0.35\textwidth}
  \centering
        \includegraphics[width = 1.0 \textwidth]{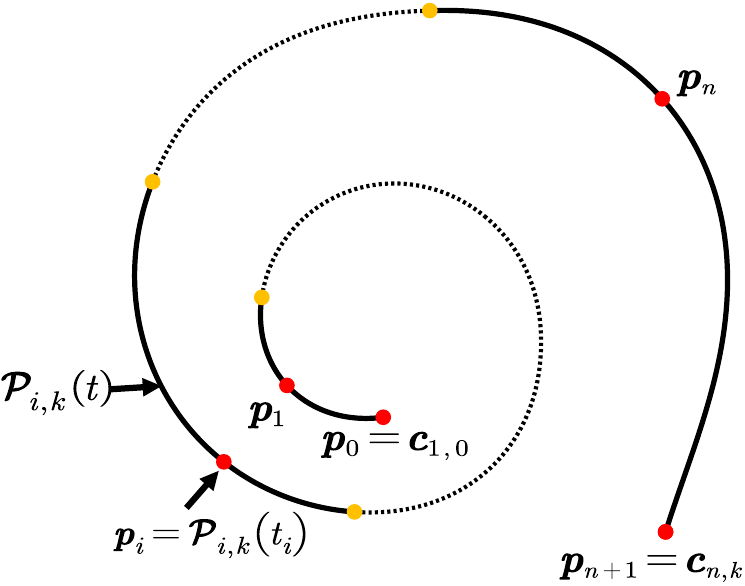}\caption{}
  \end{subfigure} \qquad
    \begin{subfigure}[b]{0.35\textwidth}
  \centering
        \includegraphics[width = 1.0 \textwidth]{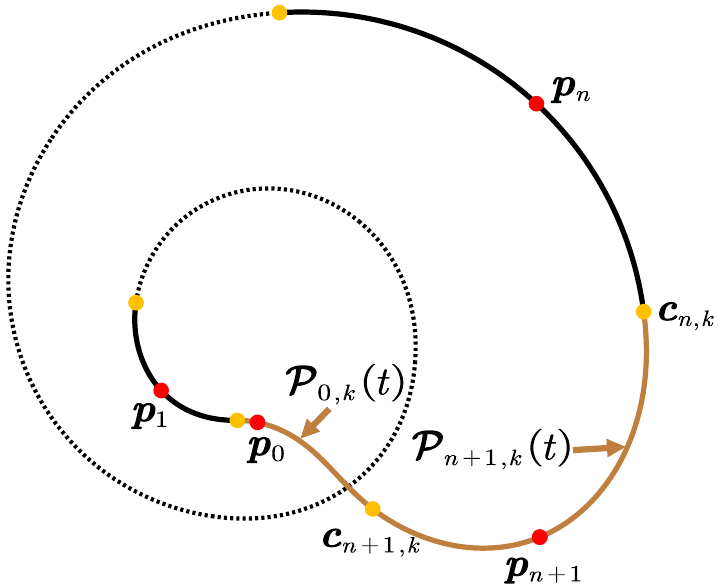}\caption{}
  \end{subfigure}
  \caption{Constraints for constructing interpolatory curves. (a) Open curve satisfying end condition $C^E$ and interpolation condition $C^I_i$, $i=1,\cdots,n$, and (b) closed curve with two additional adjacent segments, $\boldcal{P}_{0,k}$ and $\boldcal{P}_{n+1,k}$ (marked in brown), satisfying interpolation condition $C^I_i$, $i=0,\cdots, n+1$. The red and yellow points represent the interpolated points and joints of B\'ezier segments, respectively.\label{fig:constrains}}
\end{figure}

We start by discussing the constraints for constructing an open interpolatory curve. Given an ordered set of points $\{\mpp_i\}_{i=0}^{n+1}$ ($n\geq 1$), our objective is to create a curve composed of $n$ parametric curves $\{\boldcal{P}_{i,k}(t)\}_{i=1}^n$. The first curve segment should begin at the first interpolated point $\mpp_0$, and the last curve segment should end at the last interpolated point $\mpp_{n+1}$. We denote this end condition by $C^{E}$, i.e.,
\begin{description}
\item[$\bullet$] $C^{E}$: $\mc_{1,0}=\mpp_0$, $\mc_{n, k}=\mpp_{n+1}$.
\end{description}
Then, each curve segment  $\boldcal{P}_{i,k}$ $(1\leq i \leq n)$ is required to interpolate the corresponding point $\mpp_i$ at a suitable parameter value $t_i$; see Fig.~\ref{fig:constrains}(a). We denote this position interpolation constraints as $C^I_i$, which can be defined as follows:
\begin{description}
\item[$\bullet$] $C^I_i$: $\boldcal{P}_{i,k}(t_i)=\sum\limits_{j=0}^k \mc_{i, j}B_j^k\left(t_i\right) =\mpp_i, \quad t_i \in[0,1], \quad i=1,...,n$.
\end{description}
The selection of $t_i$ in the above constraints and its initialization will be discussed in Section~\ref{sec:object_function} and Section~\ref{sec:init}, respectively. Additionally, each adjacent curve pair must be smoothly connected by satisfying either parametric or geometric continuity constraints listed in Section~\ref{sec:continuity}. When constructing a closed interpolatory curve, two additional adjacent segments, denoted as $\boldcal{P}_{0,k}$ and $\boldcal{P}_{n+1,k}$ and positioned between $\boldcal{P}_{1,k}$ and $\boldcal{P}_{n,k}$, are included; see Fig.~\ref{fig:constrains}(b). The end condition $C^E$ is then replaced with two position interpolation constraints, namely $C^I_0$ and $C^I_{n+1}$.

Note that achieving higher-order continuity requires more control points compared to lower-order continuity. In our application, we utilize quartic (resp. quintic) B\'{e}zier curves ${\boldcal{P}_{i,4}}$ (resp. ${\boldcal{P}_{i,5}}$) for constructing $G^1$ or $C^1$ (resp. $G^2$ or $C^2$) interpolatory curves. This choice allows for the inclusion of more control points, which are necessary to satisfy the constraints for achieving the desired continuity, while also providing additional degrees of freedom to improve the fairness of the curve. Next, we introduce the energy function and employ optimization techniques to adjust the control points of $\boldcal{P}_{i,k}$.

\subsection{Objective function \label{sec:object_function}}
In the context of parametric curves, achieving a fair appearance is commonly associated with having a small monotonic interval of curvature and smooth curvature variation~\citep{mineur:1998:CAGD,farin2002curves}. Previous methods have focused on designing curves with monotonic curvature variations to achieve fairness~\citep{mineur:1998:CAGD,cao:2008:CAGD}. However, in the case of shape design using composed curves, it may not be necessary for each segment to have monotonic curvature. When considering the entire curve from a global perspective, it is unavoidable for complex geometric shapes to introduce many local curvature maxima. On the other hand, many aesthetically pleasing curves exhibit curvature plots that closely resemble a parabola~\citep{tsuchie2023reconstruction}. Inspired by this observation,
we propose a criterion for constructing a fairing composed B\'{e}zier curve, where each segment curve $\boldcal{P}_{i,k}$ closely approximates a specific parabola $\mathcal{Q}_i(t)$ in terms of curvature. The criterion encourages the curve to have at most two monotonic intervals and an extreme value in the curvature plot.

Note that, the curvature function of a B\'{e}zier curve is non-polynomial in terms of the parameter $t$. Hence, it is not possible to construct a B\'{e}zier curve whose curvature perfectly follows a parabolic function due to the inherent differences in their mathematical representations. To overcome this limitation, we employ optimization techniques to adjust the control points of $\boldcal{P}_{i,k}$ in order to approximate the desired parabolic curvature distribution. In order to achieve this objective, we formulate energy functions that consist of three different types of terms:
\begin{description}
\item[$\bullet$ Parabolic curvature term.] The first type of terms captures the discrepancy between the actual curvature of $\boldsymbol{\mathcal{P}}_{i,k}$ and a parabola $\mathcal{Q}_i(t)$, which can be expressed as follows:

\begin{equation}
   E_{i, p}(\mc_{i,0},..., \mc_{i,k},a_{i,0},a_{i,1},a_{i,2})=\int_0^1\left(\kappa_i(t)-\mathcal{Q}_i(t)\right)^2||\boldcal{P}_{i,k}^{\prime}(t)|| \mathrm{d} t,
\label{equ:Ekc}
\end{equation}
where $\mathcal{Q}_i(t)=a_{i,0} + a_{i,1}t + a_{i,2}t^2$ represents the parabola to be approximated by the curvature $\kappa_i(t)$, and $||\boldcal{P}_{i,k}^{\prime}(t)||\mathrm{d} t$ is the differential of arc length of the curve segment $\boldcal{P}_{i,k}$.

\begin{figure}
  \centering
  \begin{subfigure}[b]{0.5\textwidth}
  \centering
        \includegraphics[width = 1.0 \textwidth]{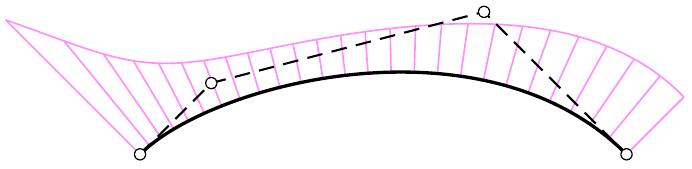}\caption{}
  \end{subfigure}
  \begin{subfigure}[b]{0.47\textwidth}
  \centering
        \includegraphics[width = 1.0 \textwidth]{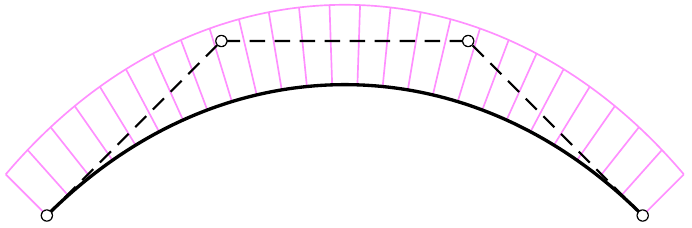}\caption{}
  \end{subfigure}
\caption{The influence of the control polygon shape on the curvature distribution of cubic B\'ezier curves. The curves in (a) and (b) have control polygons (dashed lines) with varying and similar edge lengths, respectively. The quality of the curves is visualized using the curvature comb colored in pink.}
\label{fig:extream}
\end{figure}
\item[$\bullet$ Edge length regularization term.] In fairing curve design, it is important to consider the lengths of adjacent edges of the B\'{e}zier control polygon. For instance, in typical curves~\citep{mineur:1998:CAGD,TONG2021} or Class A curves~\citep{FARIN2006,cao:2008:CAGD}, which are characterized by monotonic curvature variation, the lengths of adjacent control legs are uniformly scaled. However, when a relatively large scaling factor is employed, the curves, despite displaying a monotonic variation in curvature, might appear almost straight in certain sections. In Fig.~\ref{fig:extream}(a), we can observe that when the edge length of the control polygon in a B\'{e}zier curve varies, the curvature tends to exhibit more changes. Conversely, when the edges of the control polygon have similar lengths, the curvature varies more smoothly; see Fig.~\ref{fig:extream}(b).  Hence, the second type of terms in our energy functions restricts the variation in length between adjacent edges of a B\'{e}zier curve. This restriction is defined as follows:
\begin{equation}
E_{i, e}(\mc_{i,0},..., \mc_{i,k})=\sum_{j=0}^{k-2}\left(||\mc_{i, j} - \mc_{i, j+1}||^2-||\mc_{i, j+1} - \mc_{i, j+2}||^2\right)^2.
\end{equation}

\item[$\bullet$ Curve length term.] In our application, it is desirable to avoid curve segments with large arc lengths. Instead of directly penalizing the arc length of the curve, we address this issue by incorporating an energy term that penalizes the length of the control polygon. The definition of this energy term is as follows:
\begin{equation}
E_{i, c}(\mc_{i,0},..., \mc_{i,k})=\sum_{j=0}^{k-1}||\mc_{i, j} - \mc_{i, j+1}||^2.
\end{equation}
\end{description}

\begin{figure}
    \centering
  \begin{subfigure}[b]{0.3\textwidth}
        \centering
        \includegraphics[width = \textwidth]{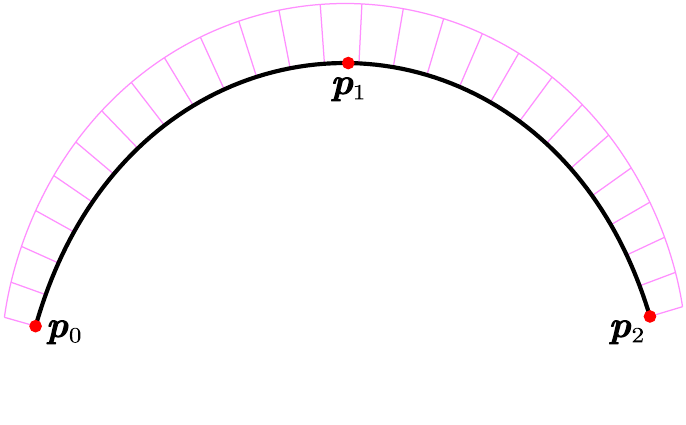}
        \caption{}
  \end{subfigure} \hspace{2.5cm}
  \begin{subfigure}[b]{0.3\textwidth}
        \centering
        \includegraphics[width = \textwidth]{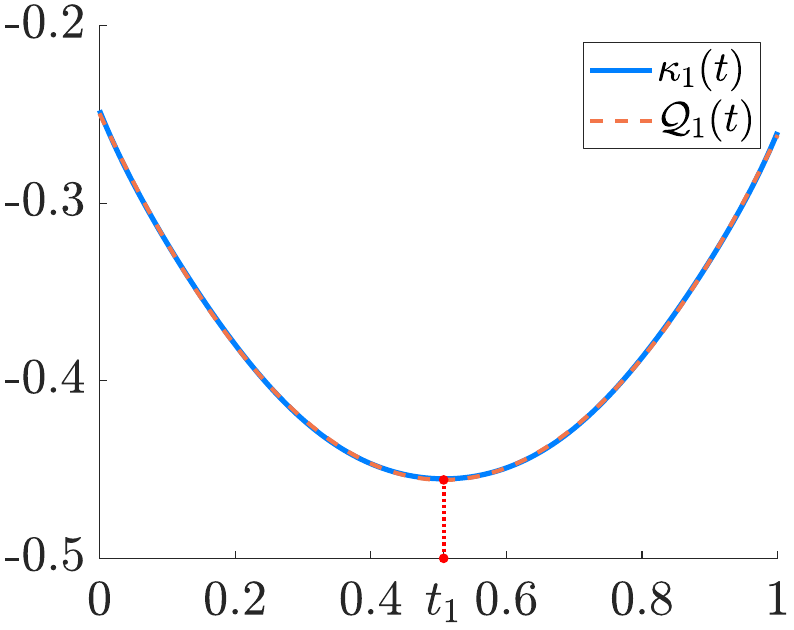}\caption{}
  \end{subfigure}
  \caption{A toy example of an open $p\kappa$-curve constructed from a single segment of quintic B\'ezier curve. (a) $p\kappa$-curve with the curvature comb (pink) interplating three points $\mpp_0,\mpp_1$ and $\mpp_2$; (b) curvature plot and appoximated parabola. The extreme value $t_1$ of the parabola is selected as the parameter for the interpolation point $\mpp_1$. }
  \label{fig:examplepkcurve}
\end{figure}

In summary, the optimization problem for constructing open $C^2$-continuous quintic $p\kappa$-curves $\{\boldcal{P}_{i,5}(t)\}_{i=1}^n$ can be formulated as follows:
\begin{equation}\label{eq:energy}
\begin{aligned}
&\min _{\mc_{i, j}, a_{i, l}} \sum\limits_{i=1}^n E_i = \min _{\mc_{i, j},a_{i, l}} \sum\limits_{i=1}^n \left(E_{i, p}+\lambda_{i, e} E_{i, e}+\lambda_{i, c} E_{i, c}\right) \\ \textrm{s.t.} &\quad \textrm{constraints} ~ C^E, \{C_{i}^I\}_{i=1}^n, ~\textrm{and}~ \{C_{i}^{C,2}\}_{i=1}^n~ \textrm{are satisfied},
\end{aligned}
\end{equation}
where $j=0,1,...,5,~l=0,1,2$, and the weights $\lambda_{i,e}$ and $\lambda_{i,c}$ are used to balance between the energy terms. The optimization problem for constructing open or closed $C^1, G^1, G^2$-continuous $p\kappa$-curves can be formulated by adapting the curve order and constraints in Eq.~(\ref{eq:energy}) to match the desired continuity and open/closed conditions. In the optimization process, the control points for each B\'{e}zier segment and the coefficients of the approximated parabola are treated as unknowns to be optimized. The parameter values of each interpolated point $\mpp_i,$ are set to $t_i = -\frac{a_{i,1}}{2a_{i,2}}$, corresponding to the parameter value at which the approximated parabola achieves extrema. In our experiments, we also enforce the constraint $t_i\in[\hat{t}_i/2, (\hat{t}_i+1)/2]$, where $\hat{t}_i$ represents the initial value assigned to the interpolation point $\mpp_i$ (see Section~\ref{sec:init}). This choice encourages the interpolation point $\mpp_i$ to be located at the curvature extrema, thereby facilitating the intuitive construction of the interpolation curve by the user, as demonstrated by Yan et al. \citep{yan2017k}. Fig.~\ref{fig:examplepkcurve} presents a toy example of a $p\kappa$-curve constructed using a single segment of quintic B\'{e}zier curve, along with its corresponding curvature distribution. As depicted in Fig.~\ref{fig:examplepkcurve}(b), when the curvature distribution of the curve closely resembles a parabola, it demonstrates a generally decreasing trend to the left of $t_i$ and an increasing trend to the right of $t_i$. The ablation study of energy terms and the discussion about the selection of weights will be deferred until Section~\ref{sec:ablation_exp}.

%
\section{Local optimization method for interactive interpolatory curve design} \label{sec:interaction}
In this section, we tackle the optimization challenge posed by the highly non-convex and non-linear problem defined in Eq.~(\ref{eq:energy}), which may involve a large number of unknowns. To overcome this issue, we propose a local optimization method that adjusts a small number of degrees of freedom at a time, allowing for efficient curve construction and editing (see Section~\ref{sec:local_construction}). Furthermore, we present an initialization strategy and a two-stage optimization approach aimed at achieving improved suboptimal local extrema in Section~\ref{sec:init} and Section~\ref{sec:two_stage}, respectively.

\subsection{Local construction method\label{sec:local_construction}}
In interactive design scenarios, the user may input or adjust one interpolation point at a time. Attempting to find the global optimal solution of the energy in Eq.~(\ref{eq:energy}) in such cases would require a global optimization from scratch, which can be computationally inefficient and may change the shape of the entire curve. To address this issue, we propose a local $p\kappa$-curve construction algorithm. When inserting a new interpolation point, it involves adding a new B\'ezier segment to the existing curve. This addition only affects specific segments of the already constructed curve.

\begin{figure}
  \centering
    \includegraphics[width = \textwidth]{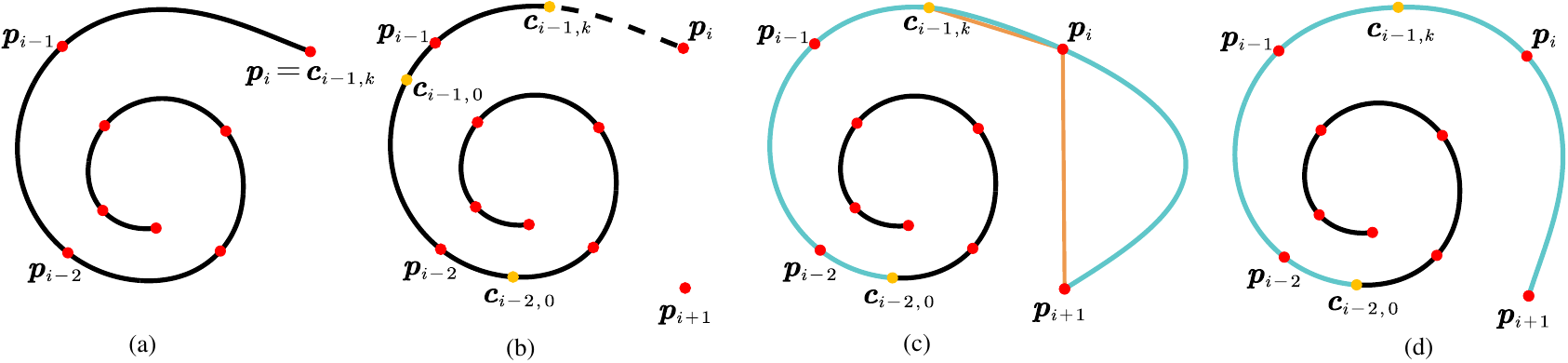}
    \caption{Local construction method for an open $p\kappa$-curve. (a) An open $p\kappa$-curve interpolating points $\{\mpp_0, \mpp_1,..., \mpp_i\}$; (b) a new point $\mpp_{i+1}$ is inserted; (c) initialization of the open $p\kappa$-curve interpolating $\{\mpp_0, \mpp_1,..., \mpp_i, \mpp_{i+1}\}$; (d) the optimized open $p\kappa$-curve. The red and yellow points represent the interpolated points and joints of B\'ezier segments, respectively. Orange polylines are used for parameter value initialization. The curve segments marked in cyan are updated when the last interpolated points are inserted. \label{fig:open_local_contruction}}
\end{figure}

\textbf{Open curve construction.} Let us discuss the construction process of the open curve as an example. {When  $\mpp_0,\cdots,\mpp_i$ ($2\leq i \leq 3$) are inserted, we solve Eq.~(\ref{eq:energy}) to obtain the corresponding curve segments.} Considering that $3 < i \leq n$ and $(i-1)$-th  B\'ezier curves have already been constructed, where $\mc_{i-1, k}=\mpp_{i}$. When the $(i+1)$-th interpolation point $\mpp_{i+1}$ is inserted, we aim to add a B\'ezier curve $\boldcal{P}_{i,k}$ that satisfies the desired geometric constraints. {In particularly, we update only the last three B\'ezier curve segments $\boldcal{P}_{i-2,k}, \boldcal{P}_{i-1,k}, \boldcal{P}_{i,k}$ that interpolate points ${\mpp_{i-2},\mpp_{i-1}, \mpp_{i},\mpp_{i+1}}$ while keeping the segments $\boldcal{P}_{1,k},..., \boldcal{P}_{i-3,k}$ fixed.}
The updated segments begin at the first control point $\mc_{i-2,0}$ of segment $\boldcal{P}_{i-2,k}$ (equally, the last control point $\mc_{i-3,k}$ of segment $\boldcal{P}_{i-3,k}$) and end at the last control point $\mc_{i, k}$ of segment $\boldcal{P}_{i,k}$. Accordingly, we solve the following simplified optimization problem:
\begin{equation}
\begin{aligned}
&\min_{\mc_{\eta, j}, a_{\eta, l}} \sum_{\eta\in\{i-2,i-1,i\}} \left(E_{\eta, p}+\lambda_{\eta, e} E_{\eta, e}+\lambda_{\eta, c} E_{\eta, c}\right) \\
\textrm{s.t.}\quad &\mc_{i,k} = \mpp_{i+1}, C_{i-2}^{C}, C_{i-1}^{C}, C_{i-2}^{I}, C_{i-1}^{I}~\textrm{and}~ C_{i}^{I}~ \textrm{are satisfied},
\label{equ:add}
\end{aligned}
\end{equation}
where the control points of the segment $\boldcal{P}_{i-2,k}$ involved in the continuity constraints $C_{i-3}^{C}$ are fixed during the optimization. Fig.~\ref{fig:open_local_contruction}(a) illustrates a toy example of an open $C^2$-continuous quintic B\'ezier interpolatory curve. When the $(i+1)$-th interpolated point is inserted, an extra B\'{e}zier segment between $\mc_{i-1,k}$ and $\mpp_{i+1}$ is introduced, which interpolates $\mpp_{i}$; see Fig.~\ref{fig:open_local_contruction}(c). {The result after optimization is shown in Fig.~\ref{fig:open_local_contruction}(d), where the three updated segments are colored in cyan.} The above process is iterated until the $n$-th B\'ezier curve $\boldcal{P}_{n,k}$ is constructed.

\begin{figure}
  \centering
    \includegraphics[width = \textwidth]{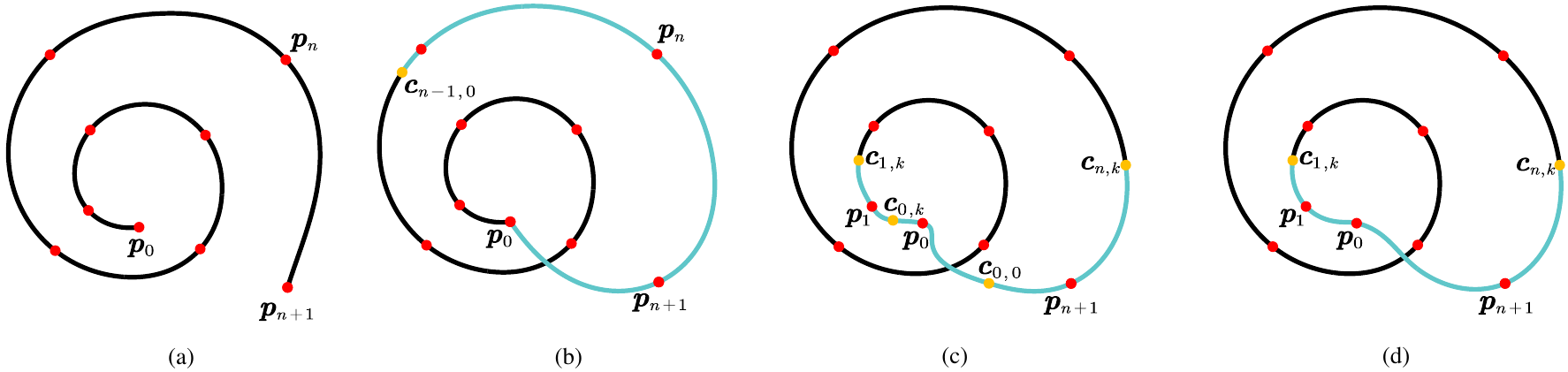}
    \caption{{Local construction method for a closed $p\kappa$-curve. (a) An open $p\kappa$-curve interpolating points $\{\mpp_0, \mpp_1,..., \mpp_{n+1}\}$; (b) a closed curve is constructed following the open curve construction method and treating $\mpp_0$ as a point inserted after $\mpp_{n+1}$; (c) initialization of the closed $p\kappa$-curve for segments $\boldcal{P}_{0,k}, \boldcal{P}_{1,k}, \boldcal{P}_{n+1,k}$ that interpolate $\mpp_0, \mpp_1$ and $\mpp_{n+1}$, respectively; (d) the optimized closed $p\kappa$-curve. The red and yellow points represent the interpolated points and joints of B\'ezier segments, respectively. The curve segments marked in cyan are updated when the open curve is closed.}} \label{fig:closed_local_contruction}
\end{figure}

\textbf{Closed curve construction.} When constructing a closed curve interpolating points $\{\mpp_0,\ldots,\mpp_{n+1}\}(n \geq 1)$, the process begins by creating an open curve following the aforementioned method. This open curve is subsequently closed to achieve the desired closed curve. Let us assume we have obtained an open curve composed of $n$ parametric curves $\{\boldcal{P}_{i,k}(t)\}_{i=1}^n$ that interpolate points $\{\mpp_0,...,\mpp_{n+1}\}$, with the first segment $\boldcal{P}_{1,k}(t)$ starting at $\mpp_0$ and the last curve segment $\boldcal{P}_{n,k}(t)$ ending at $\mpp_{n+1}$; see Fig.~\ref{fig:closed_local_contruction}(a). By straightforwardly considering $\mpp_0$ as a point inserted after $\mpp_{n+1}$ and solving the optimization problem described in Eq.~(\ref{equ:add}), a closed curve is obtained that maintains only $C^0$-continuity at the point $\mpp_{0}$; see Fig.~\ref{fig:closed_local_contruction}(b). To introduce additional degrees of freedom and achieve the desired continuity between segments, we introduce an extra segment $\boldcal{P}_{0,k}(t)$ to connect the segments $\boldcal{P}_{1,k}(t)$ and $\boldcal{P}_{n+1,k}(t)$. Then, we solve the optimization problem as
\begin{equation}
\begin{aligned}
\min _{\mc_{\eta, j}, a_{\eta, l}}  \sum_{\eta\in\{n+1, 0, 1\}}\left(E_{\eta, p}+\lambda_{\eta, e} E_{\eta, e}+\lambda_{\eta, c} E_{\eta, c}\right)
\\
\textrm{s.t.} \quad C_{n+1}^C, C_{0}^C, C_{n+1}^I, C_{0}^I~\textrm{and}~ C_{1}^I ~\textrm{are satisfied},
\label{equ:enclose}
\end{aligned}
\end{equation}
where the control points of the segments $\boldcal{P}_{n+1,k}$ and $\boldcal{P}_{1,k}$ involved in the continuity constraints $C_{n}^{C}$ and $C_{1}^{C}$ are fixed during the optimization. A toy example illustrating this process is shown in Fig.~\ref{fig:closed_local_contruction}.



\begin{figure}
    \centering
     \begin{subfigure}[b]{0.45\textwidth}
     \includegraphics[width = 0.85\textwidth]{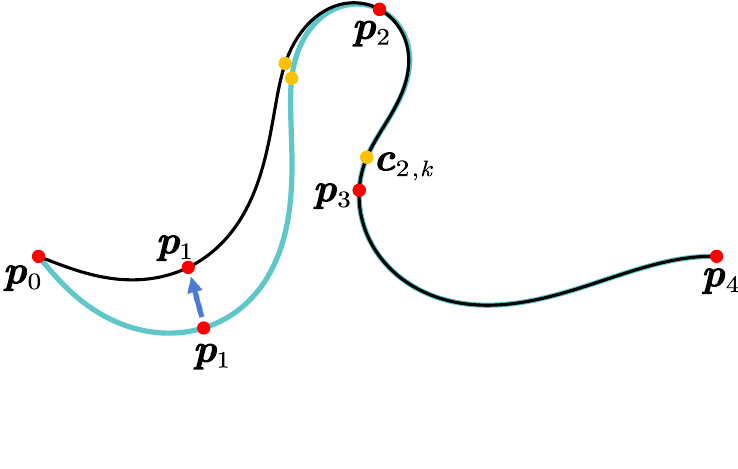} \caption{}
    \end{subfigure} \qquad
   \begin{subfigure}[b]{0.35\textwidth}
     \includegraphics[width = 0.85\textwidth]{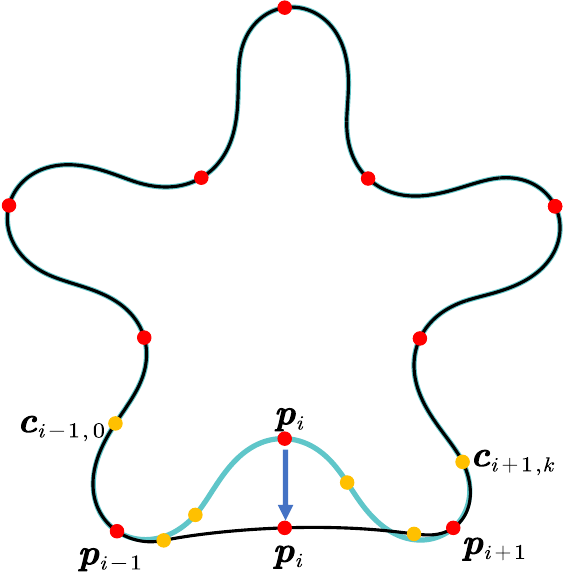} \caption{}
    \end{subfigure}
    \caption{Local Shape Editing. (a) When moving one of the first or last two interpolation points in an open $p\kappa$-curve, only two segments are affected. Otherwise, (b) relocating an interpolation point results in changes to three segments within the curve. Red and yellow points represent interpolated points and B\'{e}zier segment joints, respectively, while updated segments are colored in cyan.}
    \label{fig:move}
\end{figure}

\textbf{Local shape editing.} After the construction of the entire $p\kappa$-curve, users can make local adjustments to the shape by manipulating the position of individual interpolation points. When an interpolation point ${\mpp}_i$ is moved, only the curve segment $\boldcal{P}_{i,k}$, which interpolates ${\mpp}_i$ and its adjacent curves $\boldcal{P}_{i-1,k}$ and $\boldcal{P}_{i+1,k}$, needs to be updated. This can be achieved by solving the simplified optimization problem defined in Eq.~(\ref{equ:add}), where the involved segments with indices $\eta \in \{i-2,i-1,i\}$ are replaced by $\eta \in \{i-1,i,i+1\}$; see Fig.~\ref{fig:move}(b). Furthermore, the control points of the segment $\boldcal{P}_{i+1,k}$ that are associated with the continuity constraints $C_{i+1,k}^{C}$ remains unchanged throughout the optimization process. Similarly, for the specific cases of moving the interpolation points ${\mpp}_0$ or ${\mpp}_1$ (resp., ${\mpp}_{n}$ or ${\mpp}_{n+1}$) for an open $p\kappa$-curve, only the curves $\boldcal{P}_{1,k}$ and $\boldcal{P}_{2,k}$ (respectively, $\boldcal{P}_{n-1,k}$ and $\boldcal{P}_{n,k}$) are reconstructed (see Fig.~\ref{fig:move}(a)).

\subsection{Initialization method} \label{sec:init}
In the construction of the $p\kappa$-curve, selecting appropriate initial values for the parameter values $\hat{t}_i$ of the interpolated points $\mpp_i$, the parabolas ${\mathcal{Q}}_i$, and the curves $\boldcal{P}_{i,k}$ is crucial for the optimizing efficiency and ensuring well-behaved interpolatory curves. In the following, we provide a detailed description of the initialization process for solving the simplified problems in Eqs.~(\ref{equ:add}) and (\ref{equ:enclose}). We assume that $\mpp_i$ has been inserted and continue with the insertion of $\mpp_{i+1}$. We label the curve segments obtained after inserting $\mpp_i$ and before inserting $\mpp_{i+1}$ as $ \mathring{\boldcal{P}}_{j,k}$ $(1\leq j < i)$, where $ \mathring{t}_j$ corresponds to the parameter of interpolation point $\mpp_j$.

\subsubsection{Open curve initialization}
We will begin with the general case $i > 2$ and then address the other special case. As detailed in Section~\ref{sec:local_construction} and Fig.~\ref{fig:open_local_contruction}, when inserting $\mpp_{i+1}$, the initialization of three curve segments, namely $\boldcal{P}_{i-2,k}$, $\boldcal{P}_{i-1,k}$, and $\boldcal{P}_{i,k}$, is required. This involves determining the control points for these three curves and specifying the initialization parameters $\hat{t}_{i-2}$, $\hat{t}_{i-1}$, and $\hat{t}_{i}$ that correspond to the interpolation points $\mpp_{i-2}$, $\mpp_{i-1}$, and $\mpp_{i}$, respectively. Below, we discuss the initialization process for constructing $G^2$/$C^2$-continuity curves, noting that the initialization for $G^1$/$C^1$-continuity follows a similar approach.

\begin{description}
\item[$\bullet$ $G^2$-continuity.] {$\mathring{\boldcal{P}}_{i-2,k}$ and $\mathring{t}_{i-2}$ directly initialize the curve $\boldcal{P}_{i-2,k}$ and parameter for $\mpp_{i-2}$ when inserting $\mpp_{i+1}$, i.e., $\boldcal{P}_{i-2,k} = \mathring{\boldcal{P}}_{i-2,k}$ and $\hat{t}_{i-2} = \mathring{t}_{i-2}$.} Notably, $\mathring{\boldcal{P}}_{i-1,k}$ interpolates both $\mpp_{i-1}$ and $\mpp_{i}$. However, when inserting $\mpp_{i+1}$, $\boldcal{P}_{i-1,k}$ should be initialized to interpolate only $\mpp_{i-1}$. To achieve this, we subdivide $\mathring{\boldcal{P}}_{i-1,k}$ at the parameter value $z_{i-1} = \frac{1 + \mathring{t}_{i-1}}{2}$.  As shown in Fig.~\ref{fig:open_local_contruction}(b), the curve $\mathring{\boldcal{P}}_{i-1,k}$ in Fig.~\ref{fig:open_local_contruction}(a) is split into two curve segments, marked by a solid curve and a dashed curve, respectively. The segment $\mathring{\boldcal{P}}^{[0,z_{i-1}]}_{i-1,k}$ is used as the initialization for $\boldcal{P}_{i-1,k}$, and $\hat{t}_{i-1} =\frac{\mathring{t}_{i-1}}{z_{i-1}}$. For initializing the parameter of $\mpp_i$, we employ the chord length parameterization method, i.e., $\hat{t}_i = \frac{|\mc_{i-1,k}\mpp_i|}{|\mc_{i-1,k}\mpp_i| + |\mpp_i\mpp_{i+1}|}$ (as shown in Fig.~\ref{fig:open_local_contruction}(c)). The last segment is then uniquely initialized to achieve $C^2$-continuity with the curve segment $\boldcal{P}_{i-1,k}$ at the joint $\mc_{i-1,k}$, while also interpolating $\mpp_i$ at $t=\hat{t}_i$ and $\mpp_{i+1}$ at $t = 1$, and satisfying $\mc_{i,k-1}=\frac{\mc_{i,k-2} + \mc_{i,k}}{2}$. Fig.~\ref{fig:open_local_contruction}(c) provides a visual representation of this process for the three curve segments $\boldcal{P}_{i-2,k}$, $\boldcal{P}_{i-1,k}$, and $\boldcal{P}_{i,k}$ when inserting $\mpp_{i+1}$, all marked in cyan.

\item[$\bullet$ $C^2$-continuity.] Note that, the $G^2$-continuous initialization already achieves $G^2$-continuity at $\mc_{i-2,k} = \mc_{i-1,0}$, $C^2$-continuity at $\mc_{i-1,k} = \mc_{i,0}$, and satisfies interpolation constraints. With minor adjustments, it can serve as the initialization for $C^2$-continuous curve construction.  Specifically, we update $\mc_{i-2,k} = \mc_{i-1,0}$ and $\mc_{i-1,2}$ according to the $C^2$-continuity constraints at $\mc_{i-2,k} = \mc_{i-1,0}$ while keeping the remaining control points fixed to obtain the $C^2$-continuous initialization.
\end{description}

When only three points $\mpp_0$, $\mpp_1$, and $\mpp_2$ are inserted, we initialize a single segment $\boldcal{P}_{1,k}$ for interpolation. Specifically, we set $\hat{t}_1=\frac{||\mpp_0\mpp_1||}{||\mpp_0\mpp_1||+||\mpp_1\mpp_2||}$ and construct a quadratic B\'{e}zier curve, $\boldcal{P}_{1,2}$, to interpolate $\mpp_0$, $\mpp_1$, and $\mpp_2$ at parameters 0, $\hat{t}_1$, and 1, respectively. The curve $\boldcal{P}_{1,2}$ is then elevated to degree-$k$ and serves as the initialization for $\boldcal{P}_{1,k}$. When adding a new point $\mpp_3$, a similar method is applied to initialize $\boldcal{P}_{1,k}$ and $\boldcal{P}_{2,k}$ as the one used for initializing $\boldcal{P}_{i-1,k}$ and $\boldcal{P}_{i,k}$ when inserting $\mpp_{i+1}$.

\subsubsection{Closed curve initialization} We now address the initialization of ${\boldcal{P}_{n+1,k}, \boldcal{P}_{0,k}, \boldcal{P}_{1,k}}$ in Fig.~\ref{fig:closed_local_contruction} along with the parameter values $\hat{t}_{n+1}$, $\hat{t}_{0}$, and $\hat{t}_{1}$. Note that by employing the open curve construction method and treating $\mpp_0$ as a point inserted subsequent to $\mpp_{n+1}$, we can achieve a closed curve comprised of $n+1$ B\'{e}zier curves $\{\mathring{\boldcal{P}}_{i,k}\}_{i=1}^{n+1}$ that interpolate $\mpp_i$ at $\mathring{t}_i$, respectively.
The segments $\mathring{\boldcal{P}}_{n+1,k}$ and $\mathring{\boldcal{P}}_{1,k}$ are connected with $C^0$ continuity at $\mpp_0$.
We subdivide $\mathring{\boldcal{P}}_{n+1,k}$ and $\mathring{\boldcal{P}}_{1,k}$ at parameters $z_{n+1}=\frac{\mathring{t}_{n+1}+1}{2}$ and $z_{1}=\frac{\mathring{t}_1}{2}$, respectively. We use the chord length parameterization method and subdivision method to determine the parameters for interpolated points as follows: $\hat{t}_{n+1} = \frac{\mathring{t}_{n+1}}{z_{n+1}}$, $\hat{t}_0 = \frac{|\mc_{0,0}\mpp_0|}{|\mc_{0,0}\mpp_0| + |\mpp_0\mc_{0,k}|}$, and $\hat{t}_{1} = \frac{\mathring{t}_1}{2 - 2z_{1}}$. The initialization for ${\boldcal{P}_{n+1,k}, \boldcal{P}_{0,k}, \boldcal{P}_{1,k}}$ that maintain parametric continuity is as follows:
\begin{description}
\item[$\bullet$ $C^2$-continuity:] The endpoints of the curve $\boldcal{P}_{0,k}$ are initially set as $\mc_{0,0}=\mathring{\boldcal{P}}_{n+1,k}(z_{n+1})$ and $\mc_{0,k}=\mathring{\boldcal{P}}_{1,k}(z_1)$. Note that there are a total of 18 DOFs for the three curve segments $\boldcal{P}_{n+1,k}, \boldcal{P}_{0,k}$ and $\boldcal{P}_{1,k}$, where 3 DOFs are allocated to interpolate $\mpp_{n+1}, \mpp_{0}$, and $\mpp_{1}$ at $\hat{t}_{n+1}$, $\hat{t}_{0}$, and $\hat{t}_{1}$ respectively. Three DOFs are allocated to each endpoint $\mc_{n,k}$ and $\mc_{1,k}$, ensuring $C^2$-continuity at both joints. Meanwhile, an additional 4 DOFs are devoted to each endpoint $\mc_{0,0}$ and $\mc_{0,k}$ to attain $C^2$-continuity. To finalize the initial curves, we require the curve $\boldcal{P}_{0,k}$ to pass through the point $\frac{\mc_{0,0} + 2\mpp_{0} + \mc_{0,k}}{4}$ at parameter $t = 0.5$, using the last remaining DOF. An example of the initialization curves $\boldcal{P}_{n+1,k}, \boldcal{P}_{0,k}$, and $\boldcal{P}_{1,k}$ satisfying all constraints is shown in Fig.~\ref{fig:closed_local_contruction}(c).
\item[$\bullet$ $C^1$-continuity:] The control point $\bar{\mc}_{n+1,1}$ (resp. $\bar{\mc}_{1,k-1}$) of segment $\mathring{\boldcal{P}}^{[0,z_{n+1}]}_{n+1,k}$ (resp. $\mathring{\boldcal{P}}^{[z_{1},1]}_{1,k}$ ) is adjusted to ensure $C^1$ continuity with $\boldcal{P}_{0,k}$ at the joint $\mc_{0,0}$ (resp. $\mc_{0,k}$). This updated segment then serves as the initialization for $\boldcal{P}_{n+1,k}$ (resp. $\boldcal{P}_{1,k}$). The initialization of $\boldcal{P}_{0,k}$ is uniquely determined by the $C^1$-continuity constraints at the endpoints $\mc_{0,0}$ and $\mc_{0,k}$, along with the interpolation constraint $\boldcal{P}_{0,k}(\hat{t}_0)=\mpp_0$.
\end{description}
The initialization for constructing closed $p\kappa$-curves with $G^1$- and $G^2$-continuity is the same as that for $C^1$- and $C^2$-continuity, respectively.

\subsubsection{Parabola initialization}
For each curve $\boldcal{P}_{j,k}$ with $j=i-2,i-1,i$,  we obtain the initial values of the coefficients of the corresponding parabola $\{a_{j,2}, a_{j,1}, a_{j,0}\}$ by constrained linear least square fitting technique. Specifically, for each initial curve $\boldcal{P}_{j,k}$, the associated parabola must possess the axis of symmetry $t=\hat{t}_j=-\frac{a_{j,1}}{2a_{j,2}}$. We compute curvatures for each B\'{e}zier curve at $s$ evenly spaced points in the parameter domain. Using the constrained linear least square method, we determine the best-fitting parabola from these curvature samples, which then serves as the initialization for the parabolas. Throughout all our experiments, we set $s = 100$.


\subsection{Two-stage optimization approach} \label{sec:two_stage}
\begin{figure}
    \centering
  \begin{subfigure}{0.3\textwidth}
        \centering
        \includegraphics[width = \textwidth]{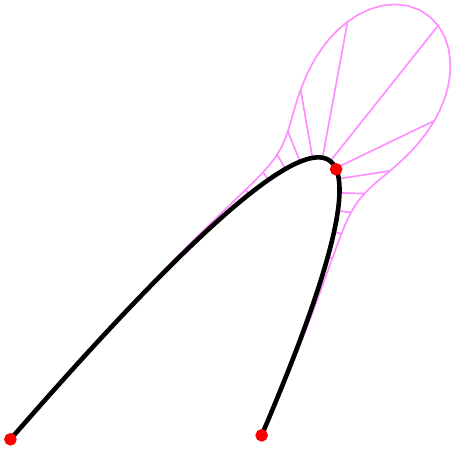} \caption{}
  \end{subfigure}
    \begin{subfigure}{0.3\textwidth}
        \centering
        \includegraphics[width = \textwidth]{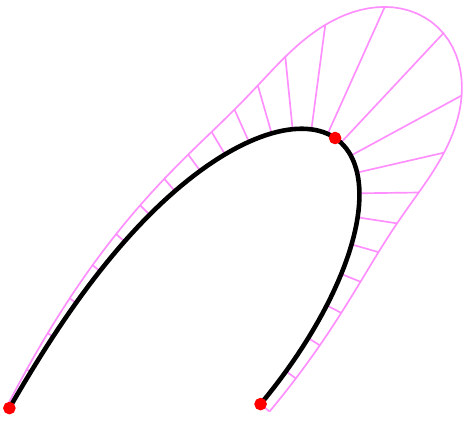} \caption{}
  \end{subfigure}
    \begin{subfigure}{0.3\textwidth}
        \centering
        \includegraphics[width = \textwidth]{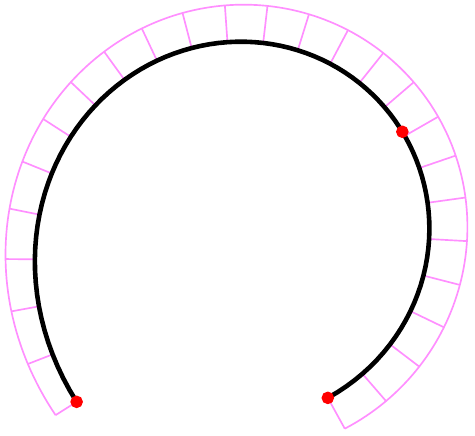} \caption{}
  \end{subfigure}
    \caption{A $p\kappa$-curve construction process based on the two-stage optimization. (a) Initial curve, and  (b\&c) results after the first and second stages of optimization. The curve quality is visualized using the curvature comb (pink), and the red points represent the interpolated points.}
    \label{fig:2step}
\end{figure}
The first stage involves formulating the objective functional using the following equation:
\begin{equation}
E_i=E_{i, p}+\lambda_{i, e} E_{i, e}+\lambda_{i, c} E_{i, c},
\label{equ:Ek}
\end{equation}
where $\lambda_{i, e}$ and $\lambda_{i, c}$ are the parameters used to penalize the energy terms $E_{i, e}$ and $E_{i, c}$, respectively. The role of the energy terms $E_{i, e}, E_{i, c}$ is further analysed in Section~\ref{sec:ablation_exp}, and the parameters $\lambda_{i, e}, \lambda_{i, c}$ are set accordingly. The first stage of optimization aims to mitigate shape changes in curvature and generates a favorable initial value for the second stage of optimization; see Fig.~\ref{fig:2step}(b). In the second stage of our process, we utilize the outcome from the first stage as the initial value for the objective functional~(\ref{equ:Ekc}), which results in the curvature function of the curve becoming closer to a parabola; see Fig.~\ref{fig:2step}(c). Both stages of optimization are terminated if the maximum value of the corresponding energy is less than the prescribed tolerance $\epsilon$ or if the number of iterations exceeds the specified number of steps $N_{iter}$. By following this process, we can ultimately achieve a satisfactory result.

\section{Experimental results and comparisons} \label{sec:experiments}
In this section, we begin by conducting ablation experiments to emphasize the significance of the energy term in Eq.~(\ref{equ:Ek}). Next, we showcase examples that are constructed using our $p\kappa$-curves. Finally, we compare our $p\kappa$-curves with state-of-the-art methods for interactive curve design. All the algorithms presented in this paper were implemented and executed using MATLAB R2023a.


\subsection{Ablation study}\label{sec:ablation_exp}
\begin{figure}
  \centering
 \begin{subfigure}{0.24\textwidth}
  \centering
\includegraphics[width = 1.0 \textwidth]{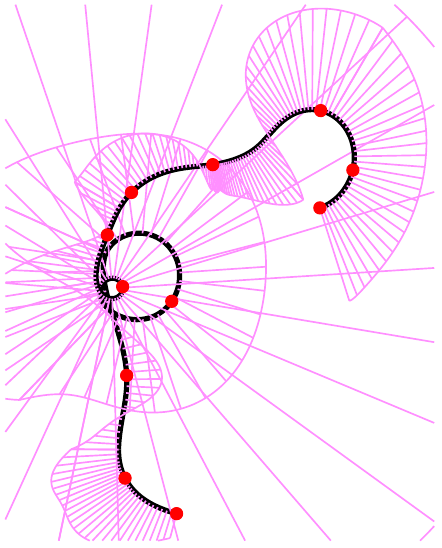}\caption{$\lambda_{i,e}=\lambda_{i,c}=0$}
  \end{subfigure}\enspace
  \begin{subfigure}{0.245\textwidth}
  \centering
\includegraphics[width = 1.0 \textwidth]{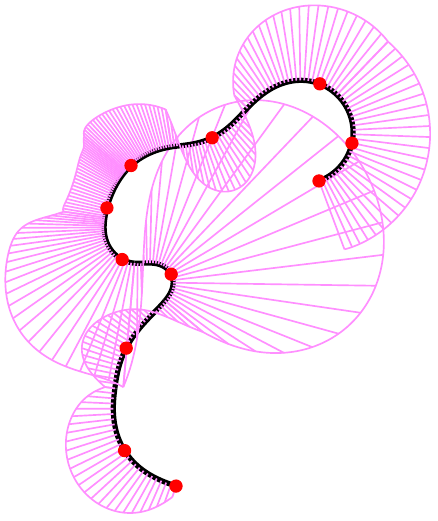}\caption{$\lambda_{i,e}=0, \lambda_{i,c}=0.1$}
\end{subfigure}\enspace
\begin{subfigure}{0.245\textwidth}
  \centering
 \includegraphics[width = 1.0 \textwidth]{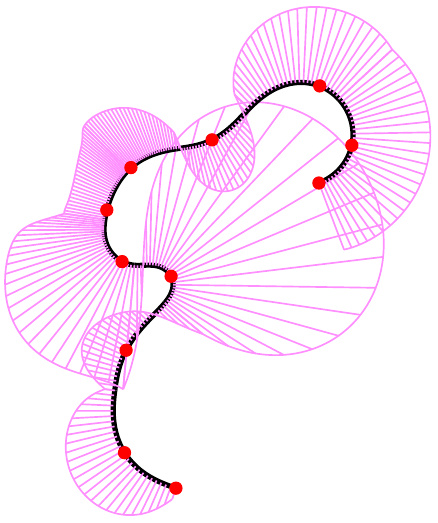}\caption{$\lambda_{i,e}=0.01, \lambda_{i,c}=0.1$}
  \end{subfigure}\enspace
    \begin{subfigure}{0.245\textwidth}
  \centering
    \includegraphics[width = 1.0 \textwidth]{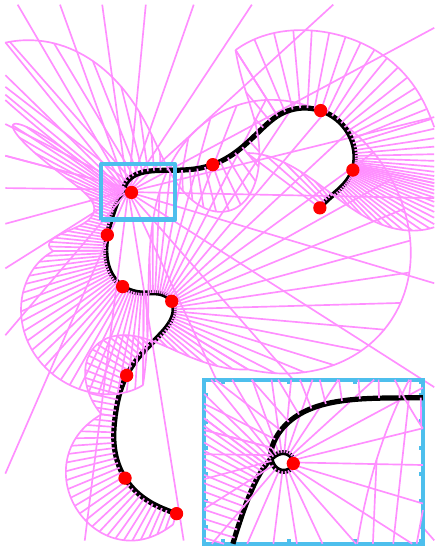}\caption{$\lambda_{i,e}=1,\lambda_{i,c}=0.1$}
  \end{subfigure} \\
    \begin{subfigure}{0.245\textwidth}
    \centering
    \includegraphics[width = 1.0 \textwidth]{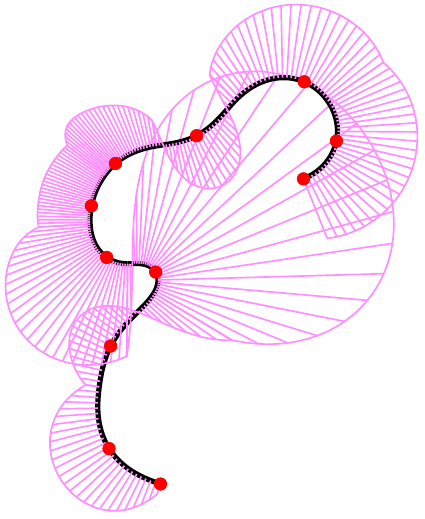}\caption{$\lambda_{i,e}=0.1,\lambda_{i,c}=0.1$}
    \end{subfigure}
    \begin{subfigure}{0.24\textwidth}
    \centering
    \includegraphics[width = 1.0 \textwidth]{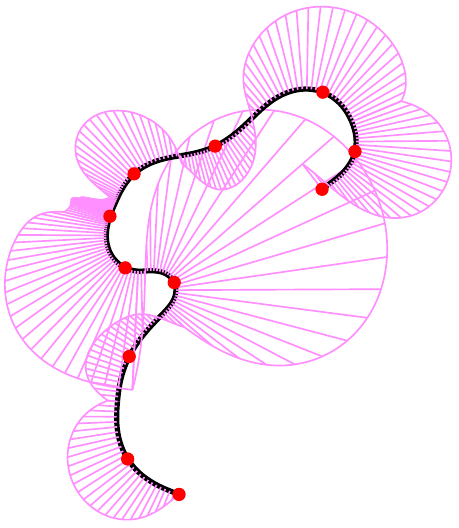}\caption{$\lambda_{i,e}=0.1,\lambda_{i,c}=0$}
  \end{subfigure}
  \begin{subfigure}{0.245\textwidth}
  \centering
  \includegraphics[width = 1.0 \textwidth]{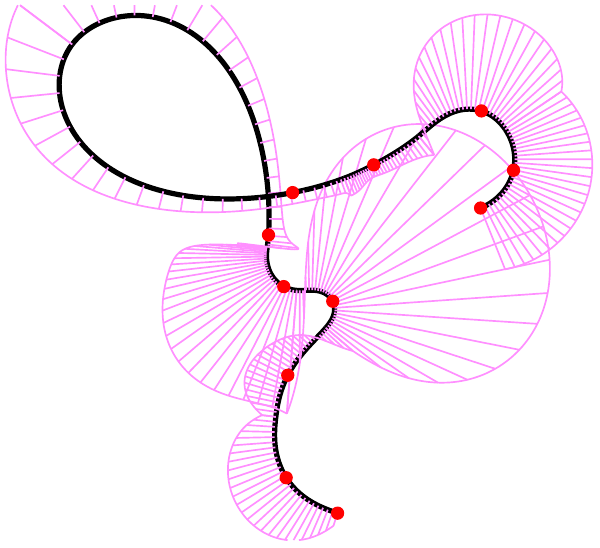}\caption{$\lambda_{i,e}=0.1,\lambda_{i,c}=0.01$}
  \end{subfigure}
  \begin{subfigure}{0.245\textwidth}
  \centering    \includegraphics[width = 1.0 \textwidth]{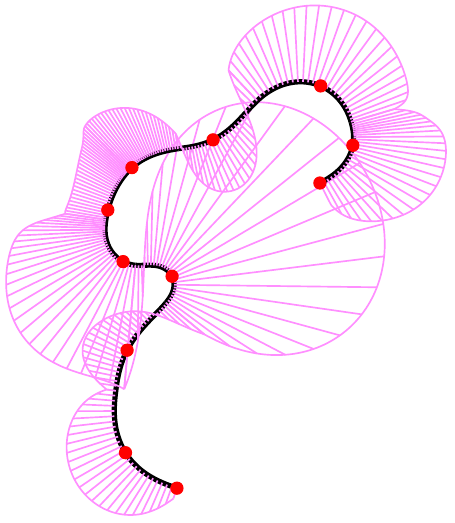}\caption{$\lambda_{i,e}=0.1,\lambda_{i,c}=1$}
  \end{subfigure}
\caption{$p\kappa$-curves for different weights $\lambda_{i,e}$ and $\lambda_{i,c}$, where red points represent the interpolated points. The curve quality is visualized using the curvature comb (pink).} \label{fig:ablation_lambda}
\end{figure}

Besides the parabola curvature term, our optimization energy defined in Eq.~(\ref{eq:energy}) also includes an edge length regularization  $E_{i,e}$ and a curve length term $E_{i,c}$. These terms penalize the difference between the lengths of adjacent edges of control points and the total curve length, respectively. In this section, we demonstrate the effect of these two energy terms and discuss the selection of corresponding weights $\lambda_{i,e}$ and $\lambda_{i,c}$. All our optimization problems are solved using the interior point algorithm~\citep{Byrd:1999:SIAM}. {{The integrals in the energy are computed by dividing the domain of integration $[0,1]$ into 100 sub-intervals and applying the composite Simpson's rule~\citep{Burden:2015} to each of these sub-intervals.}} To quantitatively evaluate the approximation of the curvature distribution to a parabola, we introduce two energy measures: the average energy $\bar{E}$ and the maximum energy $\hat{E}$ for each example. These measures are defined as follows:
\begin{equation}
 \bar{E} =\frac{1}{N}\sum\limits_{i=1}^{N} E_{i,p}, \quad\textrm{and} \quad \hat{E} = \max\{E_{i,p}|i=1,...,N\},
 \end{equation}
 where $N$ is the number of B\'{e}zier segments of the interpolatory curve.

\begin{table}
\begin{center}
\caption{Statistics of energies for different weights $(\lambda_{i,e},\lambda_{i,c})$ in Fig.~\ref{fig:ablation_lambda}. \label{tab:ablation_energy}}
\setlength{\tabcolsep}{1.5mm}{
\begin{threeparttable}
\footnotesize
\begin{tabular}{ccccccccc}
\toprule
& Fig.~\ref{fig:ablation_lambda}(a) & Fig.~\ref{fig:ablation_lambda}(b) & Fig.~\ref{fig:ablation_lambda}(c) & Fig.~\ref{fig:ablation_lambda}(d) & Fig.~\ref{fig:ablation_lambda}(e) & Fig.~\ref{fig:ablation_lambda}(f) & Fig.~\ref{fig:ablation_lambda}(g) & Fig.~\ref{fig:ablation_lambda}(h) \\
\midrule
$(\lambda_{i,e},\lambda_{i,c})$ & (0, 0) & (0, 0.1) & (0.01, 0.1) & (1, 0.1) & (0.1, 0.1) & (0.1, 0) & (0.1, 0.01) & (0.1, 1) \\
$\bar{E}$ & \num{8.33e-03} & \num{3.18e-04} & \num{3.02e-04} & \num{1.67e-02} & \textbf{\num{1.18e-04}} & \num{6.51e-04} & \num{2.27e-03} & \num{3.32e-04} \\
$\hat{E}$ & \num{4.98e-02} & \num{1.56e-03}  & \num{1.56e-03} & \num{8.36e-02} & \textbf{\num{3.24e-04}} & \num{4.07e-03} & \num{1.29e-02} & \num{1.52e-03} \\
\bottomrule
\end{tabular}
\end{threeparttable}}
\end{center}
\small{\textit{Note: The best results in the table are shown in bold font.}}
\end{table}

First, let us consider the effects of the edge length regularization term and curve length term. In the absence of these terms ($\lambda_{i,e}=\lambda_{i,c}=0$), the second optimization stage proceeds directly from the initial values. As shown in Fig.~\ref{fig:ablation_lambda}(a), the resulting curves can become excessively long and self-intersecting. This behavior occurs because, without penalties on edge lengths and overall curve length, the algorithm tends to elongate the curve to reduce curvature, leading to significant length increases. To find appropriate weights $\lambda_{i,e}$ and $\lambda_{i,c}$ for balanced energy, we first maintain $\lambda_{i,c}$ constant and vary $\lambda_{i,e}$. In Fig.~\ref{fig:ablation_lambda}(b-e), we show $p\kappa$-curves for $\lambda_{i,e}=0, 0.01, 1, 0.1$, while fixing $\lambda_{i,c}=0.1$. Results suggest that $\lambda_{i,e}\in [0,0.1]$ works well. However, larger $\lambda_{i,e}$ diminish the impact of curve length, potentially resulting in local self-intersections; see the region enclosed by the blue box in Fig.~\ref{fig:ablation_lambda}(d). Next, we consider the impact of varying $\lambda_{i,c}$ values on our resultant curves while keeping $\lambda_{i,e}$ fixed at 0.1. The corresponding outcomes are illustrated in Fig.~\ref{fig:ablation_lambda}(e-h). We note that favorable outcomes are obtained when $\lambda_{i,e}$ falls within the range of $[0,0.1]$, and $\lambda_{i,c}$ lies within the range of $[0.1,1]$, as shown in Fig.~\ref{fig:ablation_lambda}(e\&h). In Table~\ref{tab:ablation_energy}, we provide statistics for the corresponding average energy $\bar{E}$ and maximum energy $\hat{E}$ with different weight configurations. Given the lower energy value obtained in Fig.~\ref{fig:ablation_lambda}(e), we consistently set $\lambda_{i,e}=\lambda_{i,c}=0.1$ across all our examples. This choice well balances the energy terms and yields desirable curve shapes.

\begin{figure}
  \begin{subfigure}[b]{0.19\textwidth}
  \centering
        \includegraphics[width = \textwidth]{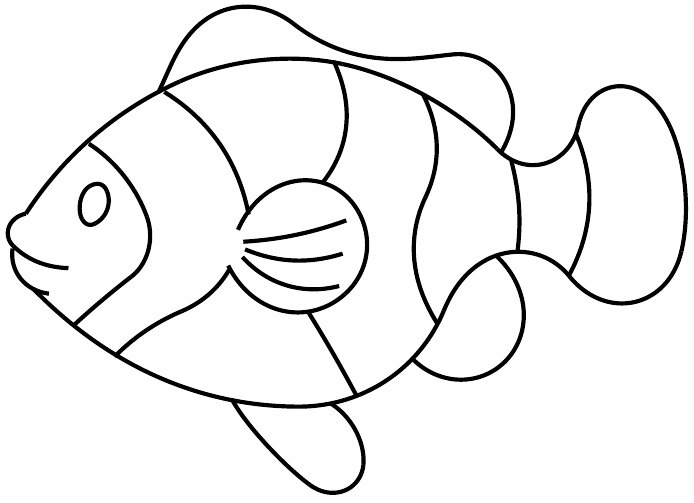}\caption*{}
  \label{fig:fish1}
  \end{subfigure}
    \begin{subfigure}[b]{0.12\textwidth}
    \centering
        \includegraphics[width =\textwidth]{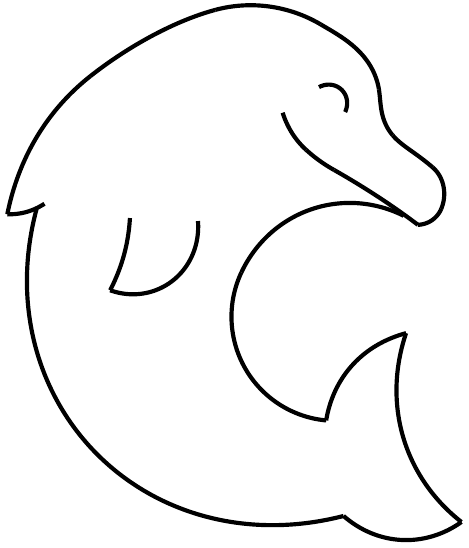}\caption*{}
  \label{fig:dolphin1}
  \end{subfigure}
    \begin{subfigure}[b]{0.17\textwidth}
  \centering
        \includegraphics[width = \textwidth]{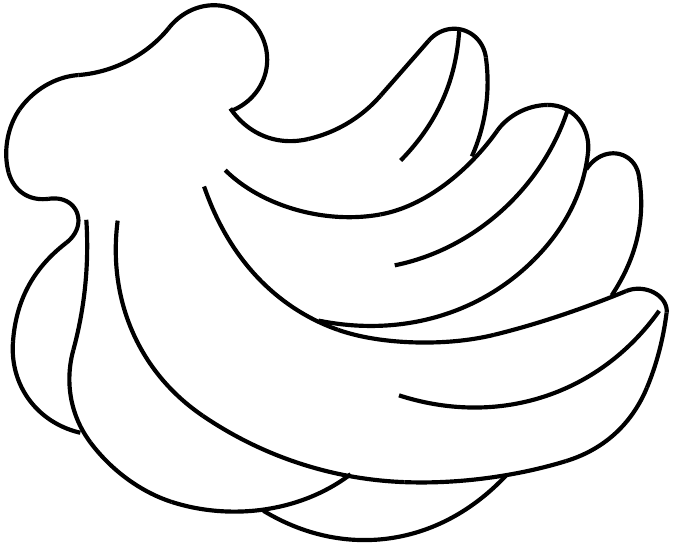}\caption*{}
  \label{fig:banana1}
  \end{subfigure}
  \begin{subfigure}[b]{0.16\textwidth}
    \centering
        \includegraphics[width = \textwidth]{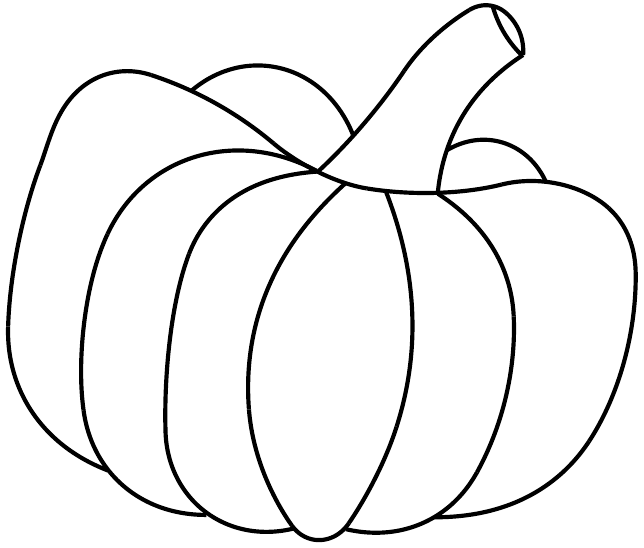}\caption*{}
  \label{fig:pumpkin1}
  \end{subfigure}
     \begin{subfigure}[b]{0.14\textwidth}
   \centering
        \includegraphics[width = \textwidth]{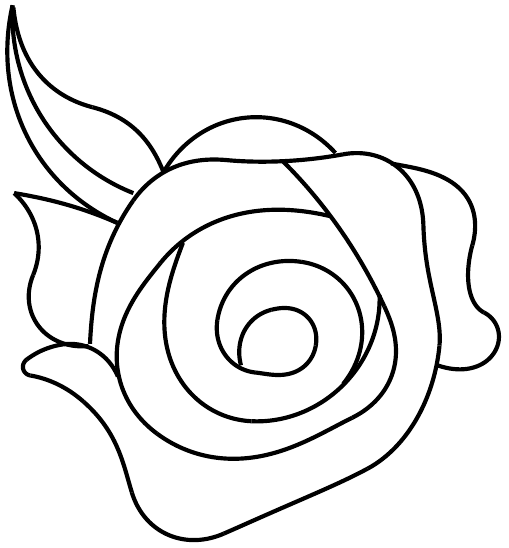}\caption*{}
  \label{fig:flower1}
  \end{subfigure}
  \begin{subfigure}[b]{0.16\textwidth}
  \centering
        \includegraphics[width = \textwidth]{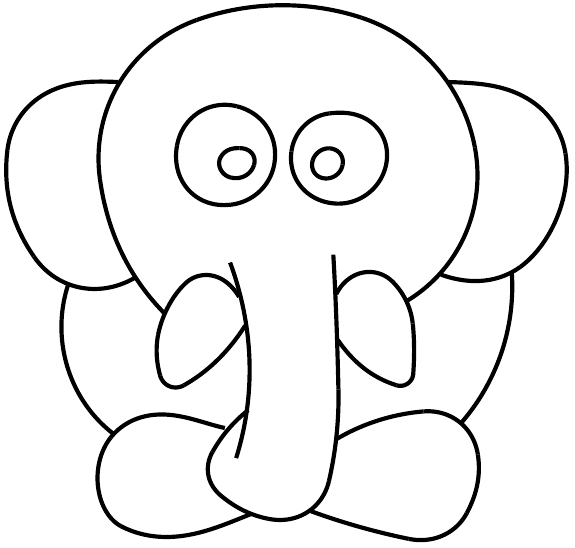}\caption*{}
  \label{fig:littleelephant1}
  \end{subfigure}
  \\
    \begin{subfigure}[b]{0.19\textwidth}
  \centering
        \includegraphics[width = \textwidth]{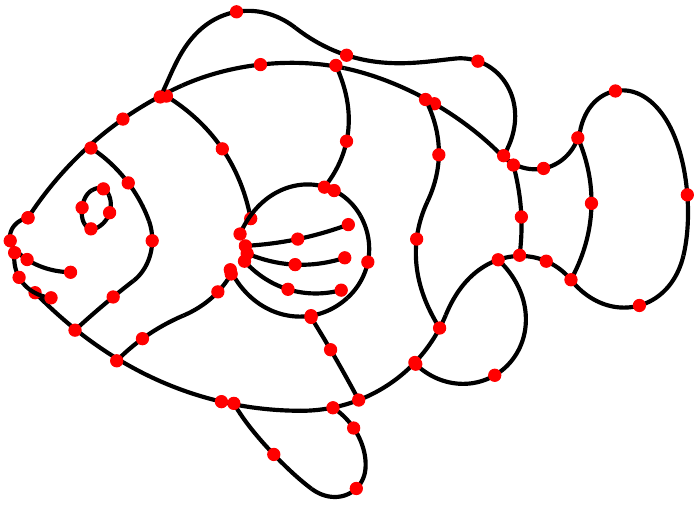}\caption{Fish}
  \label{fig:fish2}
  \end{subfigure}
    \begin{subfigure}[b]{0.12\textwidth}
    \centering
        \includegraphics[width =\textwidth]{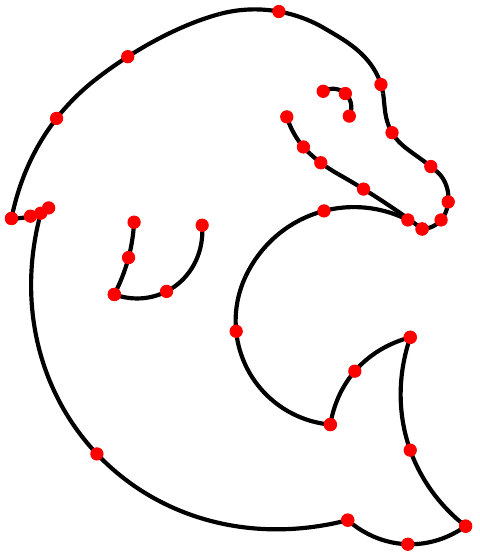}\caption{Dolphin}
  \label{fig:dolphin2}
  \end{subfigure}
    \begin{subfigure}[b]{0.17\textwidth}
  \centering
        \includegraphics[width = \textwidth]{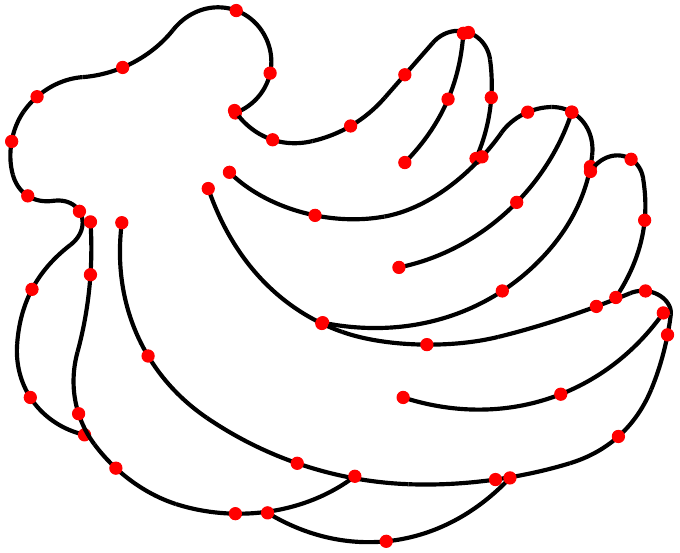}\caption{Banana}
  \label{fig:banana2}
  \end{subfigure}
  \begin{subfigure}[b]{0.16\textwidth}
    \centering
        \includegraphics[width = \textwidth]{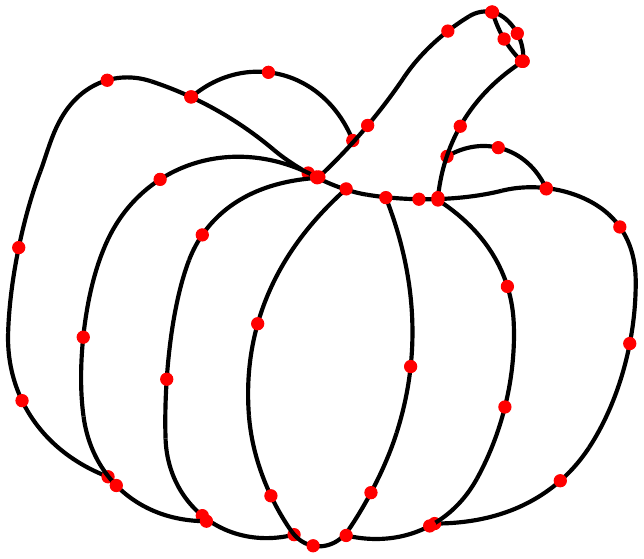}\caption{Pumpkin}
  \label{fig:pumpkin2}
  \end{subfigure}
     \begin{subfigure}[b]{0.14\textwidth}
   \centering
        \includegraphics[width = \textwidth]{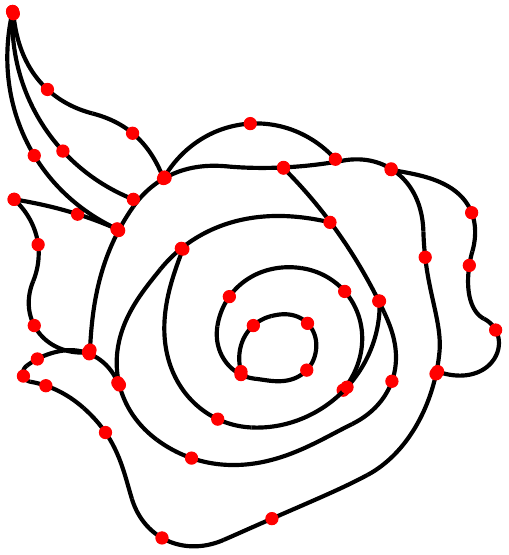}\caption{Flower}
  \label{fig:flower2}
  \end{subfigure}
  \begin{subfigure}[b]{0.16\textwidth}
  \centering
        \includegraphics[width = \textwidth]{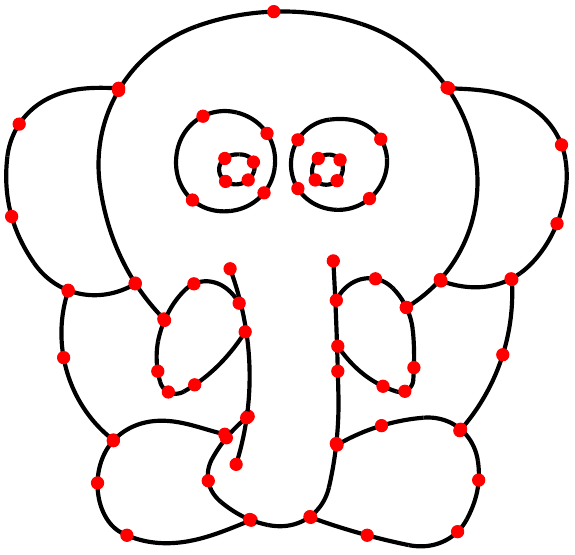}\caption{Elephant}
  \label{fig:littleelephant2}
  \end{subfigure}
\caption{Cartoon graph constructed from $p\kappa$-curves, where red points represent the interpolation points. \label{fig:cartoon}}
\begin{subfigure}[b]{0.48\textwidth}
        \includegraphics[width = \textwidth]{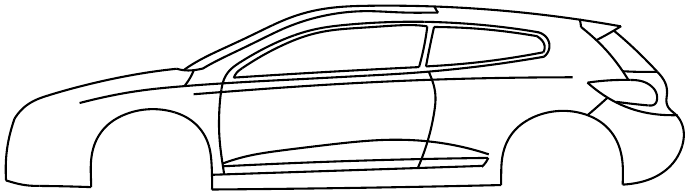}\caption{}
  \label{fig:littleelephant}
  \end{subfigure} \quad
  \begin{subfigure}[b]{0.48\textwidth}
        \includegraphics[width = \textwidth]{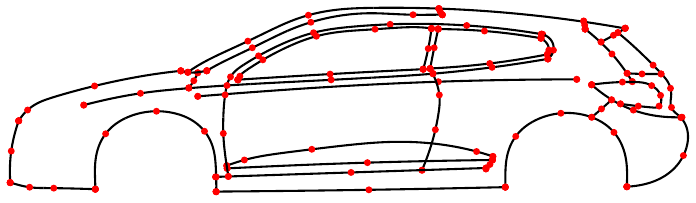}\caption{}
  \label{fig:littleelephant}
  \end{subfigure}
\caption{A car model is designed using $p\kappa$-curves, where red points represent the interpolation points. \label{fig:car}}
 \begin{subfigure}{0.275\textwidth}
     \includegraphics[width=1.00 \textwidth]{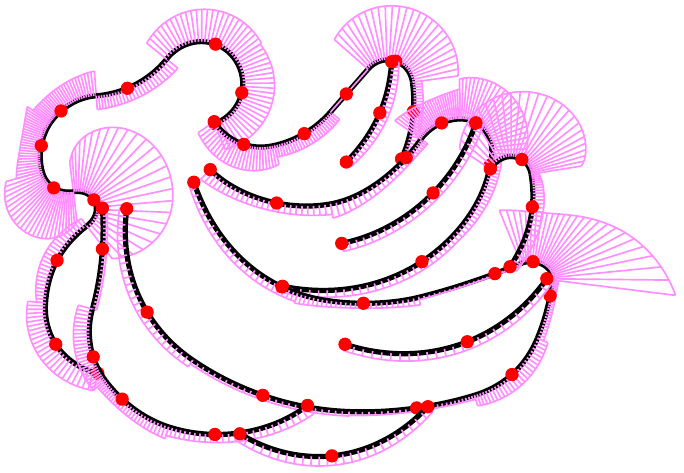}   \caption{}
  \end{subfigure}
  \begin{subfigure}{0.235\textwidth}
     \includegraphics[width = 1.00 \textwidth]{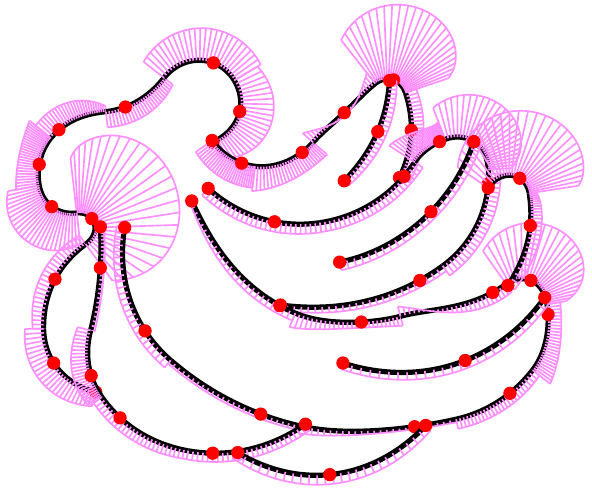}   \caption{}
  \end{subfigure}
  \begin{subfigure}{0.235\textwidth}
     \includegraphics[width = 1.00 \textwidth]{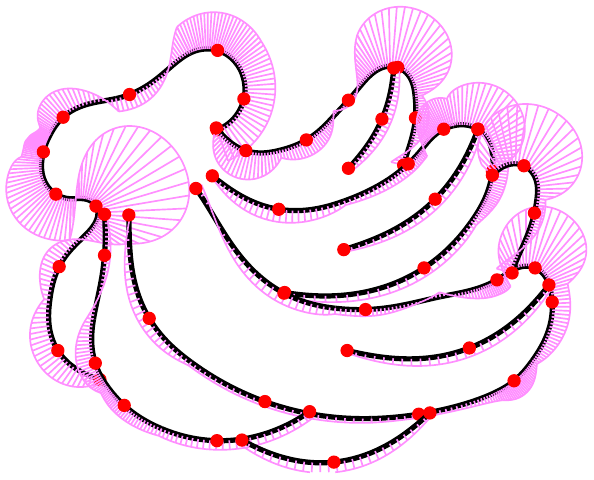}   \caption{}
  \end{subfigure}
  \begin{subfigure}{0.235\textwidth}
     \includegraphics[width = 1.00 \textwidth]{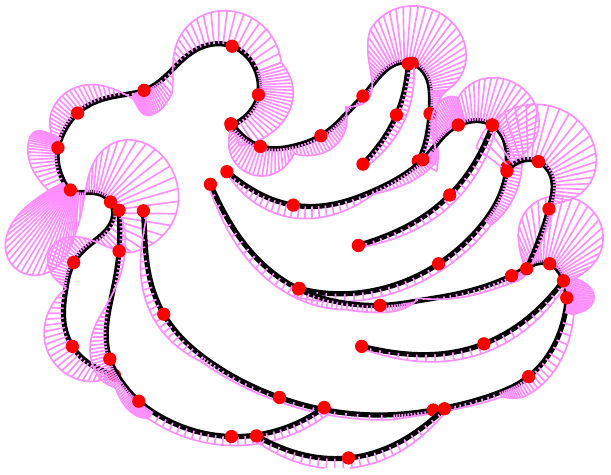}   \caption{}
  \end{subfigure}
\caption{$p\kappa$-curves with different continuity, where red points represent the interpolation points. (a) $C^1$-continuity~($\bar{E}= \num{6.17e-06}, \hat{E} =\num{1.78e-04} $); (b) $G^1$-continuity~($\bar{E}=\num{1.64e-06}, \hat{E} =\num{4.50e-05}  $); (c) $C^2$-continuity ($\bar{E}=\num{6.24e-05}, \hat{E} = \num{1.12e-3}$); (d) $G^2$-continuity ($\bar{E}= \num{3.23e-05}, \hat{E} =\num{7.75e-04} $).}
\label{fig:compare_continuity}
\end{figure}

\begin{figure}
\centering
\includegraphics[width = 1.00 \textwidth]{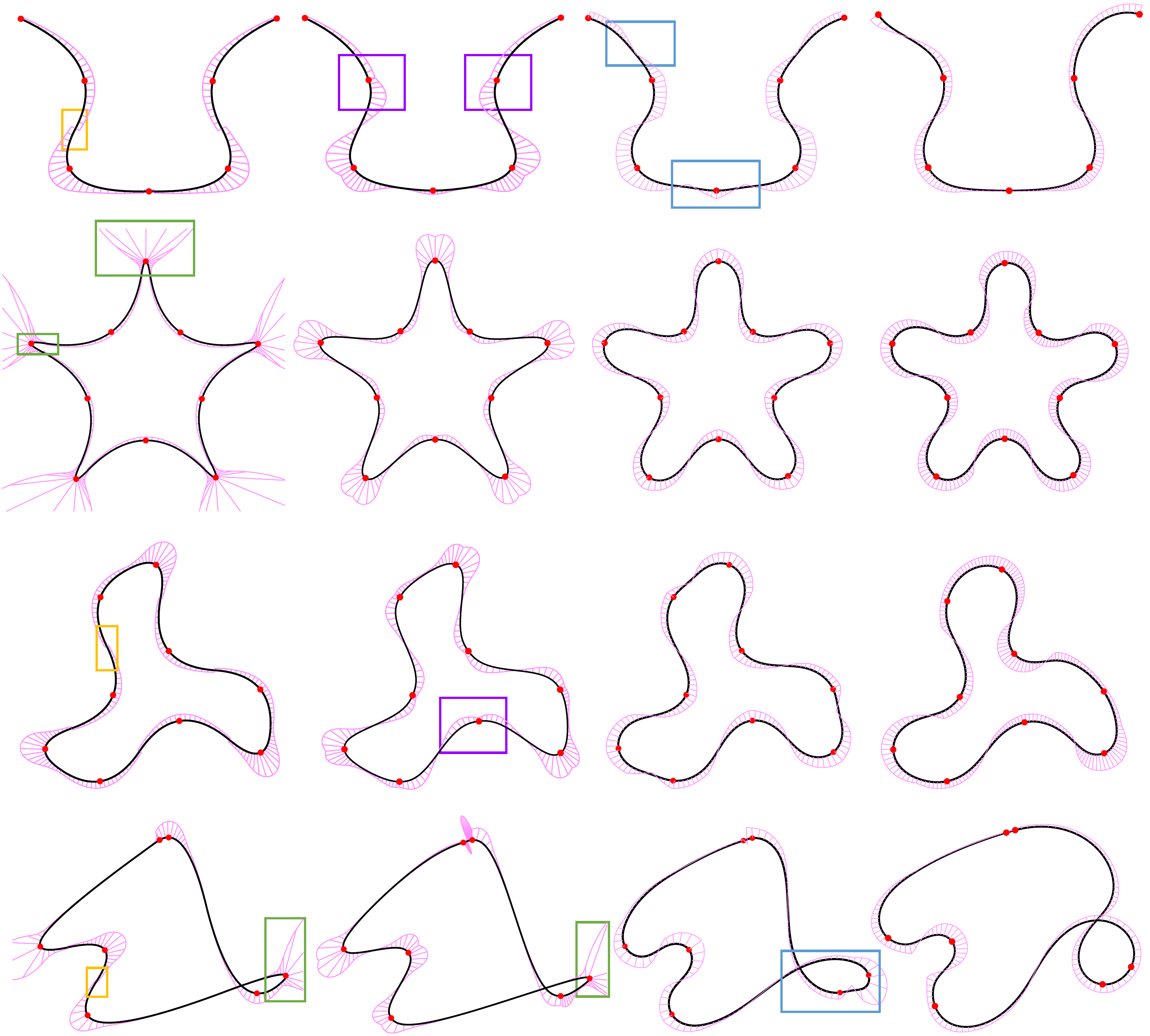}
 \begin{subfigure}{0.245\textwidth}
        \caption{}
  \end{subfigure}
  \begin{subfigure}{0.245\textwidth}
        \caption{}
  \end{subfigure}
  \begin{subfigure}{0.245\textwidth}
        \caption{}
  \end{subfigure}
  \begin{subfigure}{0.245\textwidth}
        \caption{}
  \end{subfigure}
\caption{Comparison of $p\kappa$-curves with other types of interpolation curves, where red points represent the interpolation points. (a) $\kappa$-curves~\citep{yan2017k}; (b) trigonometric blending curves~\citep{yuksel2020class}; (c) 3-arcs clothoids~\citep{binninger2022smooth};(d) $C^2$-continuous $p\kappa$-curves. Yellow and green boxes indicate curvature discontinuities and abrupt changes, respectively. Purple boxes enclose interpolation points that are not located at curvature extremes, while blue boxes enclose segments with curvatures exhibiting multiple monotonic intervals.\label{fig:compare}}
\end{figure}
\subsection{Results}

In this section, we demonstrate the application of $p\kappa$-curves in cartoon drawing and CAD sketching. Fig.~\ref{fig:cartoon} illustrates a cartoon graph constructed using multiple quartic $p\kappa$-curves with $C^1$-continuity. The interpolation points used for constructing the curves are presented in the second row. Fig.~\ref{fig:cartoon}(d-f) are specifically designed with reference to the literature~\citep{yan2017k}. Fig.~\ref{fig:car} showcases a car profile drawing created using quintic $C^2$ $p\kappa$-curves. The design of the car profile is based on the work~\citep{havemann2013curvature}. It is evident that $p\kappa$-curves consistently produce a visually fair and aesthetically pleasing result. Our method allows for the generation of $p\kappa$-curves with different geometric and parametric continuity orders. This can be achieved by specifying the desired curve order and corresponding constraints in the optimization problem. Fig.~\ref{fig:compare_continuity} depicts curves with different continuity orders generated for the same interpolated point sequence. The curvature comb is used to better visualize the differences between the results. The statistics of all the examples shown in Figs.~\ref{fig:cartoon}, \ref{fig:car} and \ref{fig:compare} are summarized in
 Table~\ref{tab:data}. The average energy $\bar{E}$ for all the curves is less than ${\num{2.00e-3}}$. The average computation time for inserting a single point is less than 0.83 seconds. Furthermore, curves with geometric continuity generally have more DOFs compared to curves with the same order of parametric continuity. This increased flexibility often enables the achievement of lower energy values in the optimization process.

\subsection{Comparisons}
In this section, we compare our curves with three state-of-the-art interpolation curves, namely $\kappa$-curves~\citep{yan2017k}, trigonometric blending curves~\citep{yuksel2020class}, and 3-arcs clothoid~\citep{binninger2022smooth}, in terms of continuity, locality, and fairness.

\textbf{Continuity.} As demonstrated earlier, our method enables the generation of $C^1$-, $C^2$-, $G^1$-, and $G^2$-continuous $p\kappa$-curves. The $\kappa$-curve, which is also composed of B\'ezier curves, achieves only $G^1$-continuity. This is illustrated in Fig.~\ref{fig:compare}(a), where the curvature of the $\kappa$-curve exhibits discontinuities at the inflection points; see the area enclosed by the yellow box. While trigonometric blending curves and 3-arcs clothoids can achieve $C^2$-continuity, it is important to note their limitations. Trigonometric blending curves use non-polynomial blend functions involving trigonometric functions, which can be computationally expensive and impractical for traditional CAD systems. Similarly, the representation of 3-arcs clothoids relies on Fresnel integrals, which are transcendental functions and may not be easily integrated into CAD systems. In contrast, our $p\kappa$-curves provide a polynomial-based representation, making them more suitable for practical CAD applications.

\textbf{Fairness.} The $\kappa$-curves and trigonometric blending curves are constructed by splicing and blending quadratic B\'ezier curves, respectively. Quadratic B\'ezier curves have limited degrees of freedom, which can make it challenging to control the variation of curvature along the curve, leading to abrupt changes in curvature and undesired visual appearance, as highlighted in the green boxes in Fig.~\ref{fig:compare}(a). $p\kappa$-curves exhibit more uniform curvature variation, avoiding abrupt changes in curvature and resulting in visually smoother curves.
Moreover, for trigonometric blending curves, we observe that the interpolated points are generally not located at the curvature extremes, as highlighted by the purple boxes in Fig.~\ref{fig:compare}(b). For 3-arcs clothoids, there are multiple curve segments (highlighted by the blue boxes) with curvatures that contain multiple monotonic intervals; see Fig.~\ref{fig:compare}(c). The use of 3-arcs clothoids guarantees the monotonicity of curvature between interpolation points with opposite curvature signs. However, when there are interpolation points with the same curvature sign, it leads to more variation in the curvature and can result in an unsatisfactory shape for the curve. The curvature distribution of our $p\kappa$-curves tends to have at most two distinct monotonic intervals, which results in a more aesthetically pleasing curve shape; see Fig.~\ref{fig:compare}(d).

\begin{figure}
\centering
  \begin{subfigure}{0.245\textwidth}
        \includegraphics[width = \textwidth]{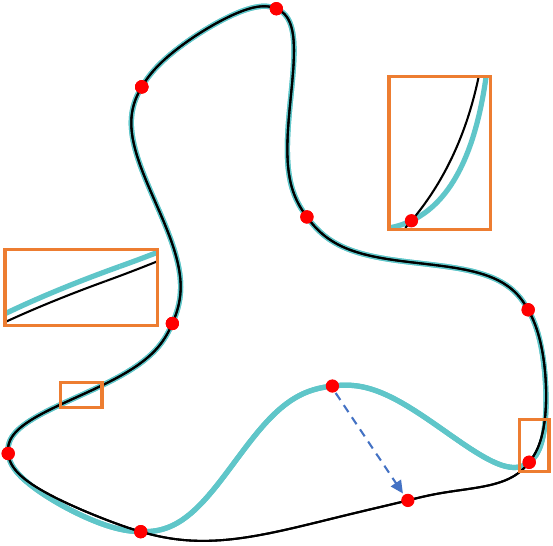}\caption{}
  \end{subfigure}
  \begin{subfigure}{0.245\textwidth}
        \includegraphics[width = \textwidth]{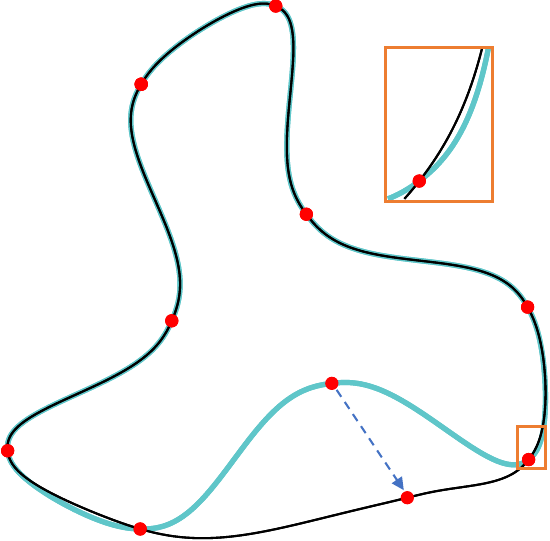}\caption{}
  \end{subfigure}
  \begin{subfigure}{0.24\textwidth}
        \includegraphics[width = \textwidth]{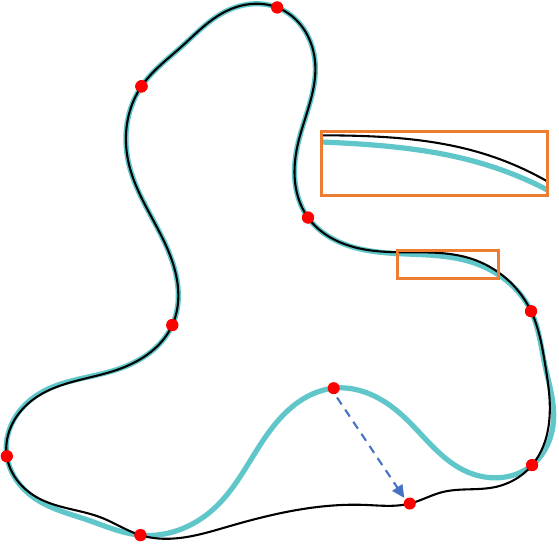}\caption{}
  \end{subfigure}
  \begin{subfigure}{0.245\textwidth}
        \includegraphics[width = \textwidth]{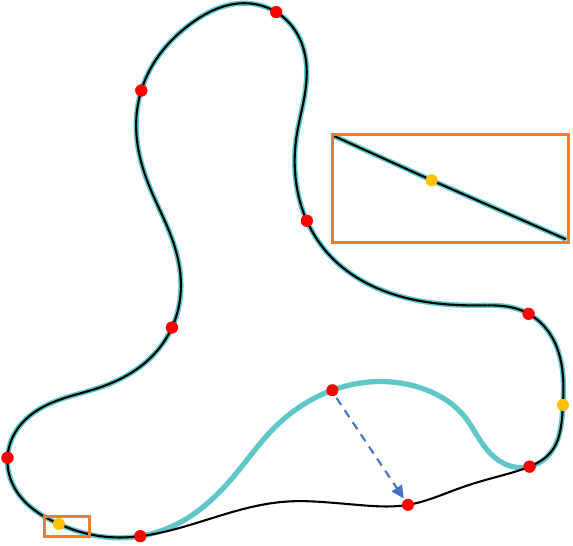}\caption{}
  \end{subfigure}
\caption{Comparison of locality when moving one interpolation point, where red points represent the interpolation points. The updated segments are colored in cyan. (a) $\kappa$-curves~\citep{yan2017k}; (b) trigonometric blending curves~\citep{yuksel2020class}; (c) 3-arcs clothoids~\citep{binninger2022smooth}; (d) $p\kappa$-curve.}
\label{fig:local}
\end{figure}

\textbf{Locality.} In Fig.~\ref{fig:local}, we compare the influence of moving one interpolating point when using different curves to interpolate the same sequence of points. Since $\kappa$-curves rely on global optimization methods, there is no guarantee that they possess locality~\citep{yan2017k}. In our experiments, we find that the influence of moving an interpolation point on the $\kappa$-curve is obvious on the segments between at least six neighboring interpolated points around the interpolation point; see Fig.~\ref{fig:local}(a). In the case of trigonometric blending curves~\citep{yuksel2020class}, each curve is obtained by blending two arcs that interpolate between three consecutive points. As a result, when an interpolation point changes position, it affects the shape of the portion between the five neighboring interpolated points; see Fig.~\ref{fig:local}(b). On the other hand, for 3-arcs clothoids, when the position of an interpolation point changes, the portion between the 6 or 8 neighboring interpolated points is affected, depending on the chosen curvature estimation method~\citep{binninger2022smooth}; see Fig.~\ref{fig:local}(c). In contrast, when changing the position of an interpolation point, our curves only undergo a shape change in at most three B\'ezier segments that interpolate the two nearest neighbor interpolated points; see the segment between two joints marked with yellow points in Fig.~\ref{fig:local}(d).

\begin{table}
\begin{center}
\caption{Statistics of $p\kappa$-curve for examples in Fig.~\ref{fig:cartoon}, Fig.~\ref{fig:car} and Fig.~\ref{fig:compare}}
\begin{threeparttable}
\begin{tabular}{l>{\centering\arraybackslash}p{2cm}c>{\centering\arraybackslash}p{2cm}c>{\centering\arraybackslash}p{2cm}cccc}
\toprule
\multicolumn{1}{l}{Example} & $N_I$ & $N_O$ & $N_C$ & $\bar{E}$ & $\hat{E}$ & $T (sec.)$   \\
\midrule
Fig.~\ref{fig:cartoon}(a)                   & 78    & 18   & 1     & \num{8.36e-05} & \num{1.75e-03} & 0.11\\
Fig.~\ref{fig:cartoon}(b)                      & 43    & 11   & 0     & \num{1.78e-06}& \num{1.74e-05} & 0.15 \\
Fig.~\ref{fig:cartoon}(c)                   & 59    & 10   & 0     & \num{3.60e-04} & \num{7.08e-03} & 0.22\\
Fig.~\ref{fig:cartoon}(d)                      & 52    & 10   & 0     & \num{1.28e-06}   & \num{2.13e-05} & 0.16\\
Fig.~\ref{fig:cartoon}(e)                   & 59    & 11   & 0     & \num{1.64e-06} & \num{4.50e-05} & 0.18\\
Fig.~\ref{fig:cartoon}(f)                     & 76    & 11   & 4     & \num{4.26e-06} & \num{1.14e-04} & 0.23\\
Fig.~\ref{fig:car}                    & 141    & 35   &0     & \num{4.45e-05} & \num{9.42e-04} & 0.76\\

\midrule
Fig.~\ref{fig:compare}(first row)                   & 7    & 1   & 0     & \num{ 4.16e-04} & \num{9.63e-04} & 0.26\\
Fig.~\ref{fig:compare}(second row)                   & 10    & 0   & 1     & \num{ 1.73E-03} & \num{2.94E-03} & 0.28\\
Fig.~\ref{fig:compare}(third row)                  & 9    & 0   & 1     & \num{ 9.96e-04} & \num{3.14e-03} & 0.28\\
Fig.~\ref{fig:compare}(fourth row)                   & 7    & 0   & 1     & \num{ 1.81e-03} & \num{4.99e-03}  & 0.83\\
\bottomrule
\end{tabular}
\small{\textit{Note: $N_I$ denotes the number of interpolated points, $N_O$ denotes the number of open curves, $N_C$ denotes the number of closed curves, and $T$ denotes the average time to insert a point.}}
\end{threeparttable}
\label{tab:data}
\end{center}
\end{table}


\section{Conclusion} \label{sec:conclusion}
This paper introduces a novel class of smooth interpolation curves called $p\kappa$-curves. The $p\kappa$-curves are constructed by joining multiple segments of B\'ezier curves, with each segment exhibiting a curvature distribution that approximates a parabola. $p\kappa$-curves can achieve the desired continuity orders such as $C^1, C^2, G^1$, and $G^2$ and demonstrate improved localizability. They also exhibit a more aesthetically pleasing appearance and a more uniformly distributed curvature compared to existing $G^2$-continuous curves. The interpolated points of a $p\kappa$-curve approximately correspond to the extrema of the segment's curvature, which provides designers with an intuitive mean to manipulate the shape of the curve. Several experiments have showcased the successful application of $p\kappa$-curves in designing a wide range of shapes, demonstrating their potential in both artistic and industrial design domains.

One limitation of our $p\kappa$-curves is that we cannot ensure symmetric shapes even when the input data is symmetric. This is due to the localized nature of our optimization method. While global optimization might improve results, it often comes at the expense of increased computational costs. Therefore, it is valuable to explore more advanced optimization methods that achieve a better balance between shape symmetry and computation time. On the other hand, as we use B\'ezier curves as primitives in our method, our curves are unable to accurately represent conics, which are commonly used in industries. Therefore, it is worth exploring the extension of our $p\kappa$-curves to include other geometric primitives, such as rational B\'ezier curves, to enhance the shape representation ability. Additionally, considering the 3D cases and accounting for torsion variations would further enhance the capabilities and applicability of $p\kappa$-curves.

\section*{Acknowledgments}
{The research of Juan Cao was supported by the National Natural Science Foundation of China
(No.~62272402) and the Fundamental Research Funds for the Central Universities (No.~20720220037). The research of Zhonggui Chen was supported by the National Natural Science Foundation of China (Nos.~61972327 and 62372389) and Natural Science Foundation of Fujian Province (No.~2022J01001).}

\bibliography{mybibfile}
\end{document}